\documentclass[aps,prc,preprint,superscriptaddress,showpacs,preprintnumbers]{revtex4}
\usepackage{graphicx}% Include figure files
\usepackage{dcolumn}% Align table columns on decimal point
\usepackage{longtable}
\begin{document}
\title{Study of nuclei in the vicinity of the
``Island of Inversion'' through fusion-evaporation reaction}
\author{R.~Chakrabarti}
\author{S.~Mukhopadhyay}
\altaffiliation[Current affiliation: ]{Department of Physics, Mississippi
State University, Mississippi State, MS 39762, USA.}
\author{Krishichayan}
\altaffiliation[Current affiliation: ]{Cyclotron Institute, Texas A$\&$M University, 
College Station, Texas 77843, USA}
\author{A.~Chakraborty}
\altaffiliation[Current affiliation: ]{Department of Physics, Krishnath College, 
Behrampore 742101, INDIA.}
\author{A.~Ghosh}
\author{S.~Ray}
\author{S.S.~Ghugre}
\author{A.K.~Sinha}
\affiliation{UGC-DAE Consortium for Scientific Research, Kolkata Centre, Kolkata 700098, INDIA}
\author{L.~Chaturvedi}
\affiliation{Department of Physics, Pt. Ravishankar University, Raipur 492010, INDIA}
\author{A.Y.~Deo}
\author{I.~Mazumdar}
\author{P.K.~Joshi}
\author{R.~Palit}
\affiliation{Tata Institute of Fundamental Research, Mumbai 400005, INDIA} 
\author{Z.~Naik}
\affiliation{Tata Institute of Fundamental Research, Mumbai 400005, INDIA}
\author{S.~Kumar}
\affiliation{Department of Physics and Astrophysics, University of Delhi, Delhi 110007, INDIA}
\author{N.~Madhavan}
\author{R.P.~Singh}
\author{S.~Muralithar}
\affiliation{Inter University Accelerator Centre, Aruna Asaf Ali Marg, 
New Delhi 110067, INDIA} 
\author{B.K.~Yogi}
\affiliation{Department of Physics, Govt. College, Kota 324009, INDIA}
\author{U.~Garg}
\affiliation{Department of Physics, University of Notre Dame, Notre Dame, IN
46556, USA}

\date{\today}

\begin{abstract}
We report the first observation of high-spin states
in nuclei in the vicinity of the ``island of inversion'', populated
via the $^{18}$O+$^{18}$O fusion reaction at an 
incident beam energy of 34 MeV. The fusion reaction mechanism
circumvents the limitations of non-equilibrated
reactions used to populate these nuclei.
Detailed spin-parity measurements in these difficult to populate nuclei 
have been possible from the observed coincidence
anisotropy and the linear polarization measurements.
The spectroscopy of $^{33,34}$P and $^{33}$S
is presented in detail along with the results of calculations within the shell 
model framework.

\end{abstract}


\maketitle

\section{Introduction}
Neutron-rich nuclei are currently of great interest as they 
exhibit structural properties very different from 
nuclei near the $\beta$-stability line. Evolving shell gaps and 
disappearance of magic numbers seen in neutron-rich nuclei 
challenge the conventional shell model theory. The ``island of
inversion" comprised of neutron-rich isotopes of Mg, Na, and 
Ne with N$\sim$20 is one of the best examples of such unexpected 
structure changes observed in nuclei with large neutron-proton 
asymmetry.
Investigations into the extent of this island and the 
transition region around it 
will lead to a greater understanding of the evolution of
the structure of the atomic nucleus.\\ 
Nuclei in and around the ``island of inversion" have in general been 
studied using transfer/deep inelastic 
reaction~\cite{Fornal,Asai,Ollier,Broda,Krishi},
or $\beta$-decay~\cite{Nathan} 
or heavy ion inelastic scattering or deuteron inelastic scattering~\cite{Iwasa}. 
However such non-equilibrated reactions have certain limitations, like 
contamination from dominant fusion-evaporation channels, low production
cross-sections, low spin population and coincident emissions
from the binary partner. The above limitations can 
be circumvented to a very large degree by using fusion-evaporation reactions with a
neutron-rich target and a neutron-rich projectile.
In this paper we present the results of a spectroscopic 
investigation  of nuclei in the 
vicinity of the ``island of inversion"($^{33,34}$P and $^{33}$S), populated
via the fusion-evaporation reaction.

\section{EXPERIMENTAL METHOD}
$^{33,34}$P and $^{33}$S nuclei were populated utilizing the
$^{18}$O+$^{18}$O reaction. The $^{18}$O beam at an incident 
energy of 34 MeV was provided by the 14 UD BARC-TIFR Pelletron
facility at TIFR, Mumbai. The choice of the incident energy 
was determined by the earlier reported excitation function
measurements~\cite{Eyal} which indicate a considerable cross-section for
these nuclei at this incident energy. The neutron-rich
$^{18}$O target was prepared by heating a 50 mg/cm$^{2}$-thick
Ta foil in an atmosphere of enriched Oxygen to form
Ta$_2$O$_5$. The total $^{18}$O equivalent thickness was estimated
to be 1.6 mg/cm$^{2}$ on both sides of the Ta foil. 
The de-exciting $\gamma$ rays were
detected by an array of 7 Compton-suppressed Clover detectors
placed at $\sim$30$^{\circ}$, $\sim$60$^{\circ}$, $\sim$90$^{\circ}$,
$\sim$120$^{\circ}$ and $\sim$150$^{\circ}$ with respect to the 
beam direction in the median plane. 
An event was recorded when
at least 2 Clovers fired in coincidence. A total of $\sim$1 billion
such $\gamma$-$\gamma$ coincidences were recorded.
The data were recorded using CAMAC based data acquisition
system LAMPS~\cite{LAMPS} and analyzed using IUCSORT~\cite{nsp1,nsp2,nsp3} 
and RADWARE~\cite{radford} software packages.
The data were pre-sorted to correct for any on-line drifts  
to ensure that there were relatively no gain changes between any two 
list mode data sets within the experiment and, then, were
precisely gain matched to ensure that data from each detector had a 
constant energy dispersion. 
The energy calibration was performed using radioactive
sources $^{152}$Eu and $^{133}$Ba and beam-off radioactivity data.
The data were sorted into symmetric and asymmetric $\gamma$-$\gamma$ 
matrices. The genetic correlation between the de-exciting $\gamma$ rays 
was established from
the symmetric $\gamma$-$\gamma$ matrix after
background subtraction and efficiency correction. 
The asymmetric matrices were used to assign the spin and parity
for the observed levels on the basis of  angular 
correlation  and the linear polarization measurements, as 
described in the next section.

\section{EXPERIMENTAL RESULTS}

\subsection{Determination of spins and parities}
\begin{figure}[htbp]
\includegraphics[trim=0.0cm 0.0cm 0.0cm 0.0cm, scale=0.45,angle=270]{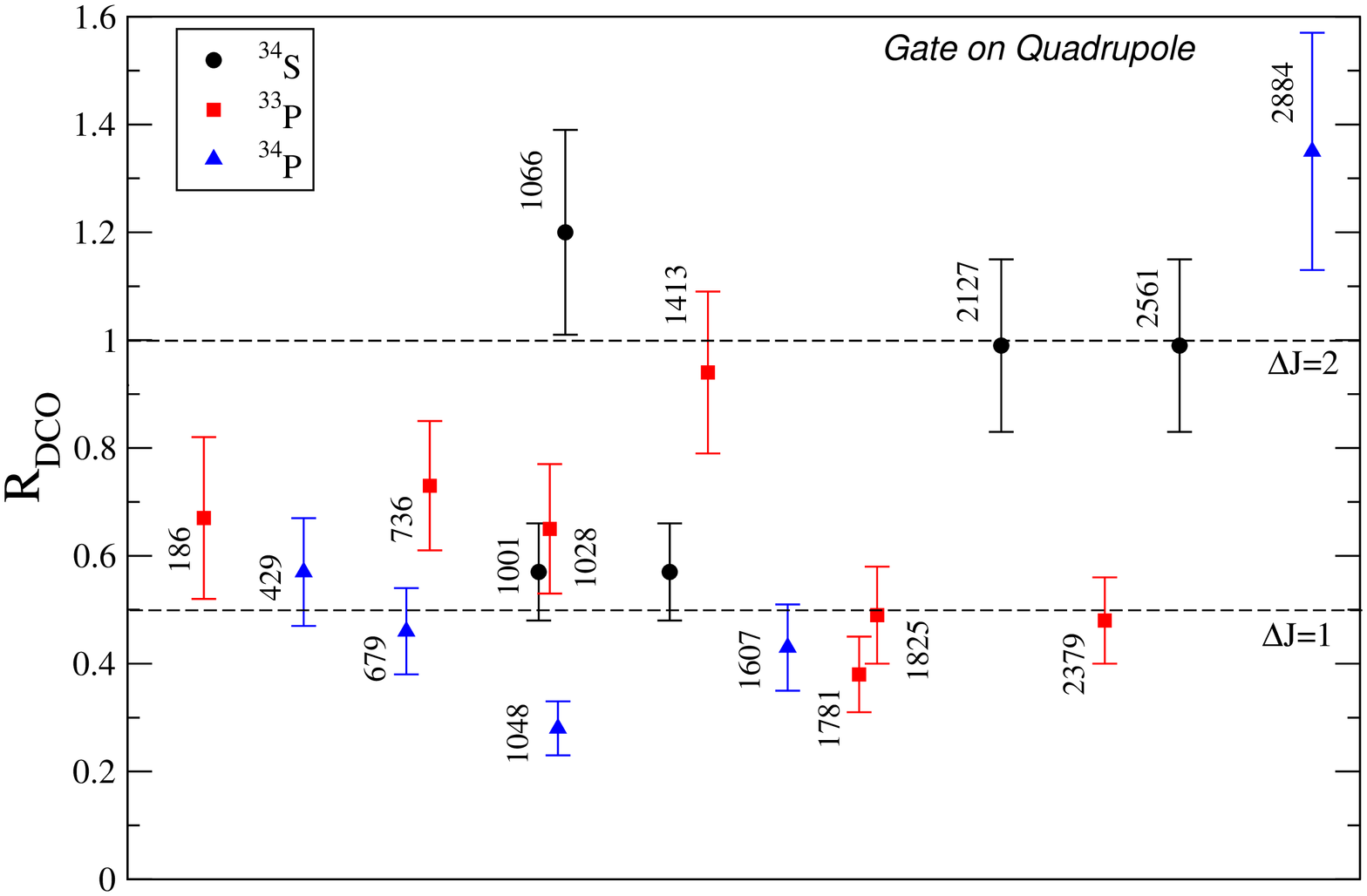}
\caption{\label{fig:RDCOq}``(Color online)" The experimental $R_{DCO}$ values for transitions
in $^{33,34}$P and $^{33,34}$S when the gate is on a quadrupole transition.}
\end{figure}

\begin{figure}[htbp]
\includegraphics[trim=0.0cm 0.0cm 0.0cm 0.0cm, scale=0.45,angle=270]{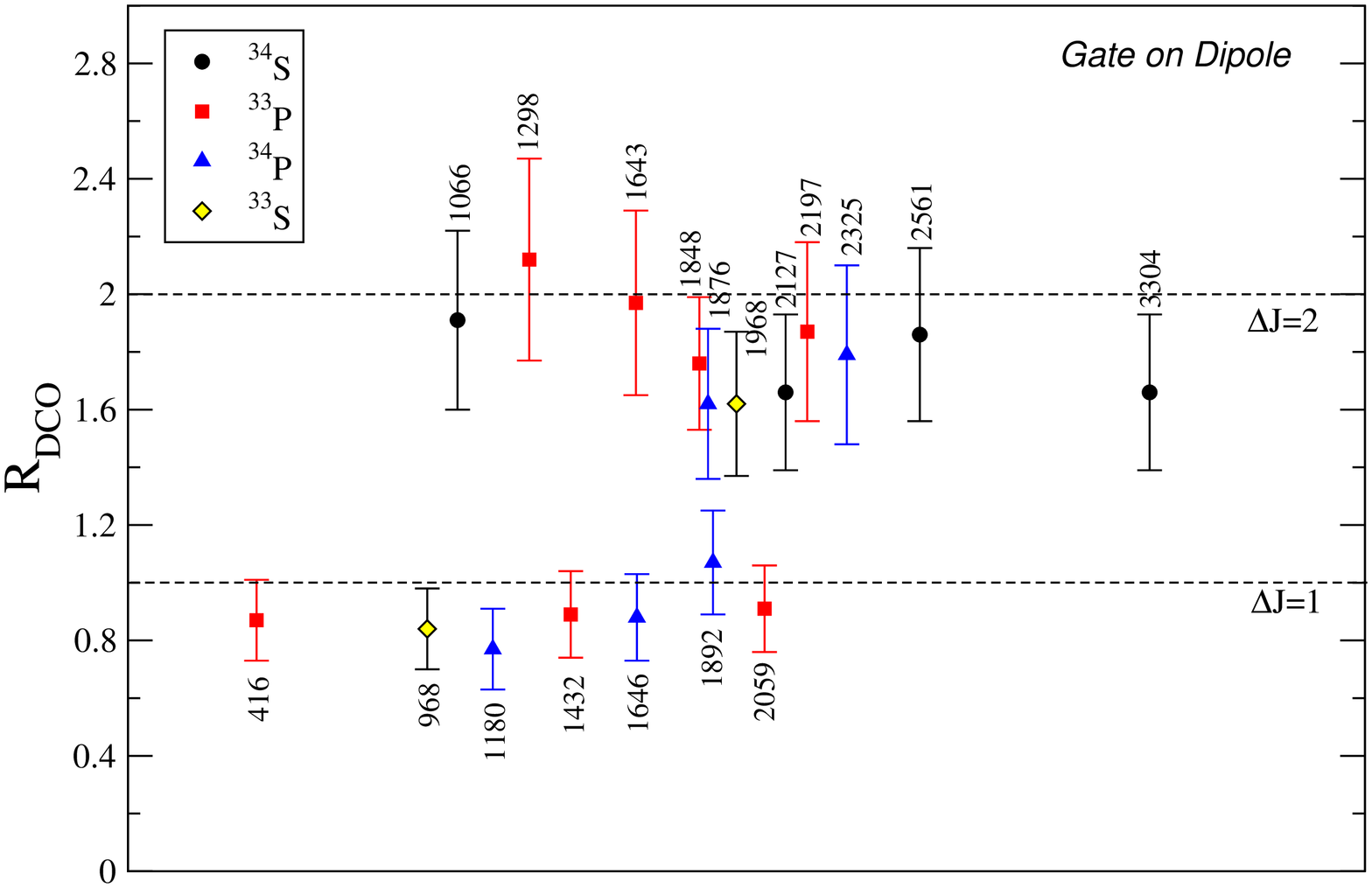}
\caption{\label{fig:RDCOd}``(Color online)" The experimental $R_{DCO}$ values for transitions
in $^{33,34}$P and $^{33,34}$S when the gate is on a quadrupole transition.}
\end{figure}

The multipolarity assignments have been performed from the observed coincidence
angular correlations. Assuming pure (stretched) transition, the
coincidence intensity anisotropy can be used to distinguish between
$\Delta$J = 1 and $\Delta$J = 2 transitions. 
A qualitative assignment for the multipolarity of the
$\gamma$-transition from the angular correlation measurements
is obtained following the procedure detailed in Ref.~\cite{Krane}.
The experimental $R_{DCO}$
in the present work is defined as:
\begin{equation}
\label{eq-1}
R_{DCO} = \frac {I_{\gamma1} (at~\theta~gated~by~\gamma_{2}~at~90^{0})}
{I_{\gamma1} (at~90^{0}~gated~by~\gamma_{2}~at~\theta)}
\end{equation}

where $\theta$ is 30$^{0}$ and 150$^{0}$.
When the gating transition ($\gamma_{2}$) is a stretched quadrupole transition, 
$R_{DCO}$ $\sim$1 for a pure quadrupole ($\gamma_{1}$) and $\sim$0.5 for stretched
pure dipole transition($\gamma_{1}$). Similarly, a gate on a dipole transition would result in
$R_{DCO}$ $\sim$2 for a pure quadrupole and $\sim$1 for a pure dipole transition.
These intensity ratios were obtained from the angle-dependent ${\gamma}$-${\gamma}$
matrices assuming stretched transition for the gates and after incorporating
necessary efficiency corrections. The experimental $R_{DCO}$ values
determined from quadrupole and dipole gates have been plotted  for several transitions
belonging to $^{33,34}$P, $^{33,34}$S in  Fig.~\ref{fig:RDCOq} and Fig.~\ref{fig:RDCOd} respectively.
The present statistics did not permit us to extend these measurements to the weak transitions.
As seen from the figures, it is possible to distinguish between $\Delta$J = 1,
and $\Delta$J = 2 transition following the above procedure. For mixed transitions
the plot essentially provides a qualitative way of determining the dominant multipolarity
considering the proximity of the $R_{DCO}$ value
to the $\Delta$J = 1 or the $\Delta$J = 2 line.

The angular correlation measurement is not sensitive to the
electric or magnetic character of the radiation. The information on this was
obtained from the linear polarization measurements. Clover detectors
have an advantage over conventional single crystal detectors 
as they allow such measurements to be made.

\begin{figure}[htbp]
\includegraphics[trim=0.0cm 0.0cm 0.0cm 0.0cm, scale=1.0,angle=0]{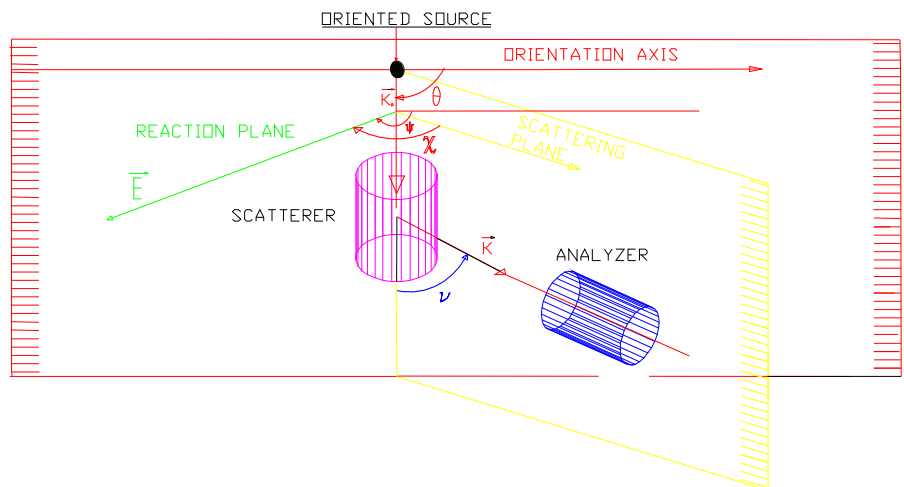}
\caption{\label{fig:pol}``(Color online)" Schematic diagram of the geometry of a Compton
polarimeter}
\end{figure}

The angular distribution of linearly polarized gamma rays from an axially
oriented ensemble of nuclei is given by~\cite{Steffan,Fagg}
\begin{equation}
\label{eq-2}
W(\theta,\psi) = \frac {d\Omega}{8\pi} \sum_{\lambda=even}
B_{\lambda}U_{\lambda} \left [A_{\lambda}P_{\lambda}(cos \theta)
+2A_{\lambda 2}P_{\lambda 2}^{(2)}(cos \theta)~cos~2\psi \right]
\end{equation}

where B$_\lambda$ are orientation tensors describing the degree of orientation
of the parent nucleus and U$_\lambda$ are deorientation coefficients. 
P$_\lambda$ are the ordinary Legendre polynomials and P$_{\lambda2}^{(2)}$ 
are the unnormalized associated Legendre polynomials. 
A$_\lambda$  are angular distribution coefficients which depend
on the spin of the initial and final state and the multipolarity
of the $\gamma$-transition. The coefficients A$_{\lambda2}$ depend
on the electromagnetic character of the radiation~\cite{Deng}.
$\theta$ is the angle that the electric vector of the emitted quanta
makes with the orientation axis and $\psi$ is the angle
between the electric vector E of the emitted quanta and the reaction 
plane (Fig.~\ref{fig:pol}).

The degree of linear
polarization P$\theta$ of a $\gamma$ ray is defined as the difference between
the intensities of the radiation presenting an electric vector parallel to
the reaction plane ($\psi$ = 0$^{0}$) and that with an electric vector perpendicular
to the plane ($\psi$ = 90$^{0}$)~\cite{Deng,Jones,Starosta}:
\begin{equation}
\label{eq-3}
P(\theta) = \frac {W(\theta, \psi=0) - W(\theta, \psi=\pi/2)}
{W(\theta, \psi=0) + W(\theta, \psi=\pi/2)}
\end{equation}

where the normalization is such that 
-1 $\leq$ P($\theta$) $\leq$ +1.\\ 
P($\theta$)= 0 for an 
unpolarized $\gamma$-ray and has a maximum value  
at $\theta$ = 90$^{0}$,

\begin{equation}
\label{eq-4}
P_{cal}(90^{0}) = \pm \frac {3a_{2}H_{2}-7.5a_{4}H_{4}}
{2-a_{2}+0.75a_{4}}
\end{equation}

where {\em a$_{2}$} and {\em a$_{4}$} are the angular distribution coefficients and
the {\em H$_{2}$} and {\em H$_{4}$} coefficients depend on the initial and final spin and the
mixing ratio, $\delta$~\cite{Aoki,Mateosian}.

Experimentally linear polarization of gamma rays was detected and measured
through Compton scattering~\cite{Deng}.
The differential Compton scattering cross-section is given by~\cite{Deng}
\begin{equation}
\label{eq-5}
\frac {d\sigma}{d\Omega}(\nu,\chi)= \frac {r_{0}^{2}}{2}
\left (\frac {E}{E_{0}}\right )^2 \left[\frac {E_{0}}{E} + \frac {E}{E_{0}}
- 2~sin^{2}\nu~cos^{2}\chi \right ]
\end{equation}

where {\em r$_{0}$} is the classical electron radius, $\nu$ is the Compton
scattering angle with respect to the direction of the incident $\gamma$ ray,
and $\chi$ is the angle between the electric vector E of the primary radiation
and the scattering plane defined by the direction of the incident and the
scattered photons (Fig.~\ref{fig:pol}).
This cross-section is relatively high and polarization sensitive for
a wide photon energy range. Maximum scattering occurs at $\chi$ = 90$^{0}$. 

The Clover detectors used in the experiment acted 
effectively as Compton polarimeters.
The detectors placed at $\sim$90$^{0}$ were particularly useful since
polarization is maximum in that direction. Each crystal of a Clover detector
acts as a scatterer and the two adjacent crystals act as the absorbers. The asymmetry
between the perpendicular and parallel scattering with respect to the reaction
plane distinguishes between electric and magnetic transitions.
The experimental asymmetry or $\Delta_{IPDCO}$ (IPDCO stands for ``Integrated Polarizational-
Directional Correlation from oriented nuclei'')
at 90$^{0}$ between perpendicular and parallel coincidence rates is defined~\cite{Starosta} as
\begin{equation}
\label{eq-6}
\Delta_{IPDCO}=\frac {aN_{\perp} - N_{\parallel}} {aN_{\perp} + N_{\parallel}}
\end{equation}

where N$_\perp$ and N$_\parallel$ are the number of photons with a given energy
scattered along the direction perpendicular and parallel to the reaction plane,
respectively, in the detectors placed at $\sim$90$^{0}$ and in coincidence
with another photon detected in at least one other detector in the array.
This is called an integrated PDCO because the polarization of one
$\gamma$ quantum is measured and the information is integrated over all the
possible emission directions of the accompanying coincident radiation.

``{\em a}" denotes the correction due to the asymmetry
in response of the clover segments. This factor is energy dependent
({\em a} = {\em a$_{0}$}+{\em a$_{1}$}{\em E$_{\gamma}$}), and is determined using a radioactive
source (having no spin alignment) under similar conditions. 
This correction is
defined as~\cite{Starosta,Jones} 

\begin{equation}
\label{eq-7}
a=\frac{N_\parallel(unpolarized)}{N_\perp(unpolarized)}
\end{equation}

The values for {\em a$_{0}$} and {\em a$_{1}$} for the present experimental
setup were 1.00007(0.00698) and
7.89667$\times$10$^{-7}$ keV$^{-1}$(5.61462$\times$10$^{-7}$)
respectively.

\begin{figure}[htp]
\includegraphics[trim=0.0cm 0.0cm 0.0cm 0.0cm,scale=0.50,angle=270]{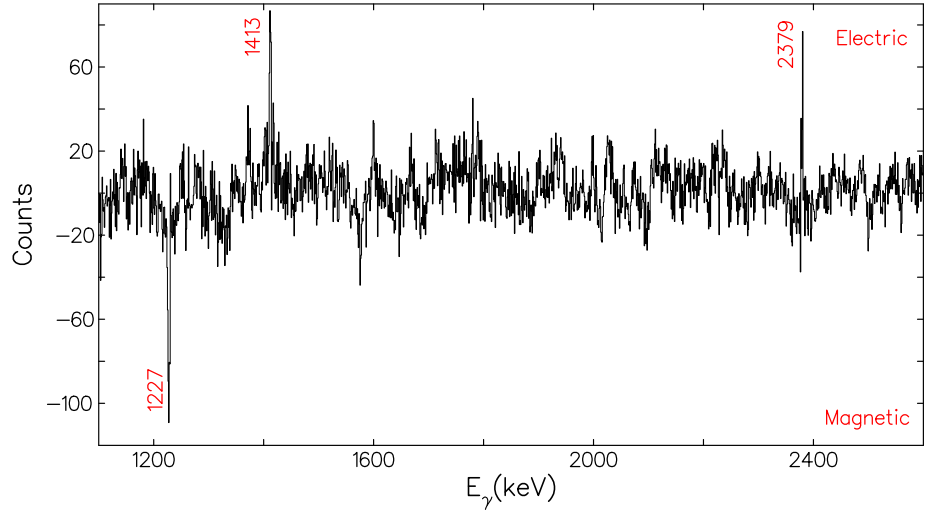}
\caption{\label{fig:4}``(Color online)" Background subtracted difference spectrum for perpendicular 
and parallel coincidences. The positive peaks indicate electric transitions
and the negative peaks indicate magnetic transition. Gate is on 1848 keV (5/2$^{+}$ $\rightarrow$ 1/2$^{+}$)
in $^{33}$P.}
\end{figure}

$\Delta_{IPDCO}$ values were evaluated from asymmetric $\gamma$-$\gamma$
matrices whose one axis corresponds to the perpendicular or parallel
scattered events in the clovers at $\sim$90$^{0}$ and the other axis 
corresponds to the total energy recorded in any of the other 
detectors.
Gates were put on the full energy peaks of the perpendicular
and parallel matrices to obtain spectra
representing either perpendicular or parallel scattering respectively,
and from these, the $N_\perp$ and $N_\parallel$ values were obtained for each 
transition. Fig.~\ref{fig:4} is a representative background subtracted difference spectrum 
of perpendicular and parallel gates. The positive peaks indicate
electric transitions whereas negative peaks indicate magnetic transitions.

The linear polarization is related to the asymmetry by the
polarization sensitivity Q(E$_{\gamma}$) as~\cite{Starosta,Jones}
\begin{equation}
\label{eq-8}
\Delta_{IPDCO} = PQ(E_{\gamma})
\end{equation}

\begin{figure}[htbp]
\includegraphics[trim=0.0cm 0.0cm 0.0cm 0.0cm, scale=0.45, angle=270]{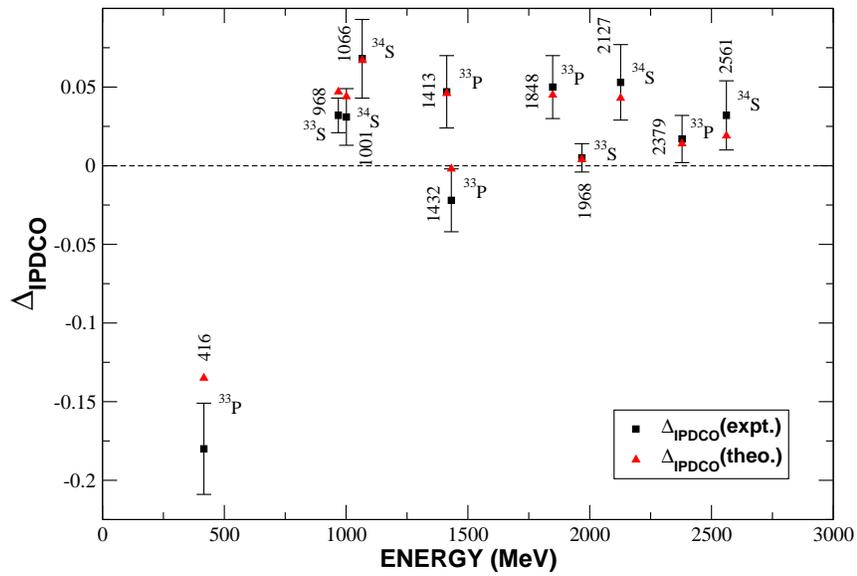}
\caption{\label{fig:IPDCO}``(Color online)" Theoretical and experimental $\Delta_{IPDCO}$
as a function of gamma ray energy.}
\end{figure}

{\em Q(E$_{\gamma}$)} is dependent on the incident gamma-ray energy and the
geometry of the polarimeter and its values, were obtained for a similar setup
as reported by Palit {\em et al.}~\cite{Palit}.
The theoretical polarizations were determined using eq.(4) by calculating the
angular distribution coefficients {\em $a_{2}$}, {\em $a_{4}$} and the {\em H$_{2}$}, {\em H$_{4}$}
functions for each value of the multipole mixing ratio $\delta$ using the formalism
given in Refs.~\cite{Aoki,Mateosian,Yamazaki}.
The theoretical $\Delta_{IPDCO}$ values were obtained from the theoretical polarization
values using eq.(8), and are plotted along with the experimentally obtained $\Delta_{IPDCO}$ values
as a function of $E_\gamma$ in Fig.~\ref{fig:IPDCO} for several strong transitions in
$^{34}$S, $^{33}$P and $^{33}$S. The results show a good agreement
between the theoretical and experimental values. At a given energy, a positive
value of the asymmetry parameter indicates an electric transition,
a negative value indicates a magnetic transition and a near-zero value is indicative
of an admixture.

\begin{figure}[htbp]
\includegraphics[trim=0.0cm 0.0cm 0.0cm 0.0cm, scale=0.45,angle=270]{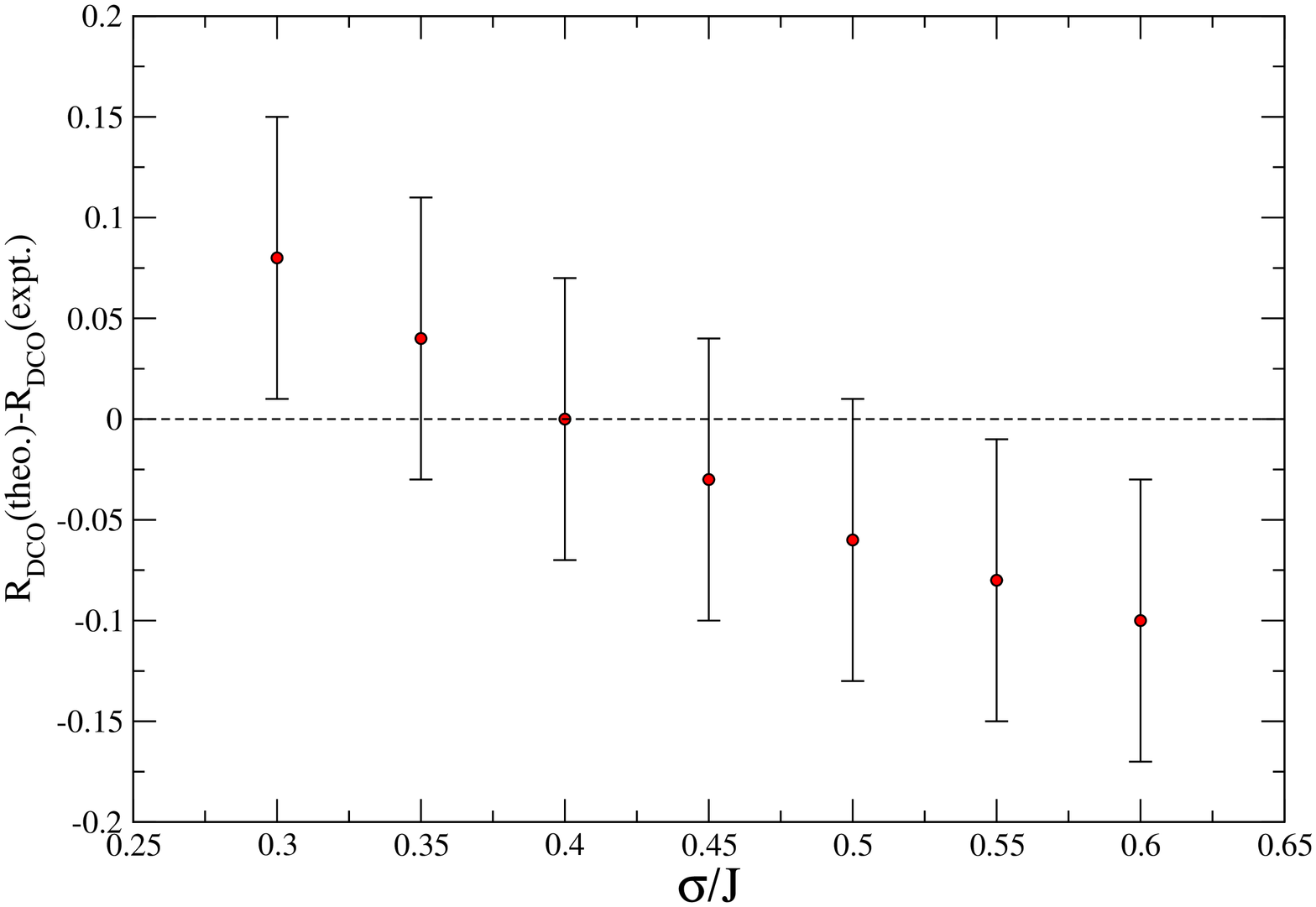}
\caption{\label{fig:6}``(Color online)" Plot of the difference between theoretical and 
experimental $R_{DCO}$ as a function of $\sigma$/J for 1066 keV
transition (5$^{-}$ $\rightarrow$ 3$^{-}$) in $^{34}$S.}
\end{figure}

\begin{figure}[htbp]
\includegraphics[trim=0.0cm 0.0cm 0.0cm 0.0cm, scale=0.45, angle=270]{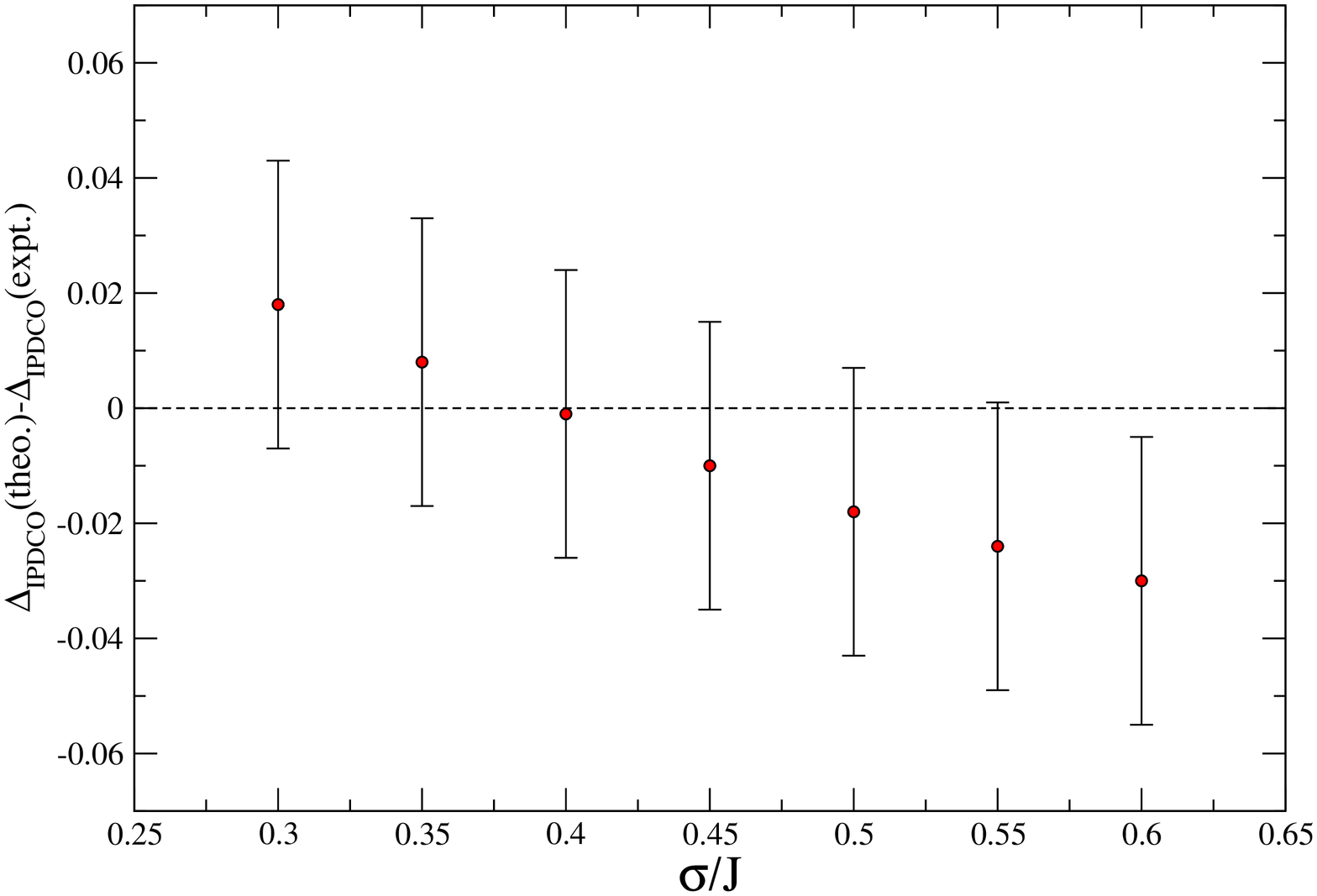}
\caption{\label{fig:7}``(Color online)" Plot of the difference between theoretical and
experimental $\Delta_{IPDCO}$ as a function of $\sigma$/J for 1066 keV 
transition (5$^{-}$ $\rightarrow$ 3$^{-}$) in $^{34}$S.}
\end{figure}

The calculation of theoretical polarization requires two inputs, {\em viz.},
the width of the {\em m-state} distribution and the mixing ratios.
Polarization depends significantly on the distribution of the nuclear state over
its magnetic substates. When the alignment is partial, the angular distribution coefficients
for complete alignment~\cite{Yamazaki} have to be multiplied with
the attenuation coefficients as formulated by 
Der Mateosian and Sunyar~\cite{Mateosian},
which depend on the factor {\em $\sigma/J$} where $\sigma$ is the width of the distribution
of the {\em m}-states, assuming a Gaussian distribution~\cite{Yamazaki}. The choice of $\sigma/J$
was made following the simultaneous analysis of the $R_{DCO}$ and $\Delta_{IPDCO}$ values
for several transitions of known spin and parity.
Fig.~\ref{fig:6} and Fig.~\ref{fig:7} depict the determination of $\sigma/J$ for
one such transition, the 1066-keV [5$^{-}$ $\rightarrow$ 3$^{-}$, $\delta$ = 0]
belonging to $^{34}$S. This nucleus has been populated with substantial cross-section in the
present experiment and its level scheme  has been extensively reported earlier
by Mason {\em et al.}~\cite{Mason}. The theoretical $R_{DCO}$ values were obtained using
the code ANGCOR~\cite{Krane2} as a function of $\sigma/J$.
Fig.~\ref{fig:6} represents the comparison of the theoretical $R_{DCO}$ values
with the experimentally obtained values. As seen from the figure, a value of $\sigma/J$
$\sim$0.35 - 0.45 appears to be reasonable. A similar plot has been made for the
$\Delta_{IPDCO}$ values of the same transition in Fig.~\ref{fig:7}. Here also the
theoretical values agree with the experimental ones in the same range.
This exercise was repeated for other transitions in $^{34}$S of known mixing ratios~\cite{NNDC}.
It was observed that a value of $\sigma/J$ = 0.4 consistently reproduced the observed
$R_{DCO}$ and $\Delta_{IPDCO}$ values and hence this value was used in all our calculations.

\begin{figure}[htbp]
\includegraphics[trim=0.0cm 0.0cm 0.0cm 0.0cm, scale=1.0,angle=0]{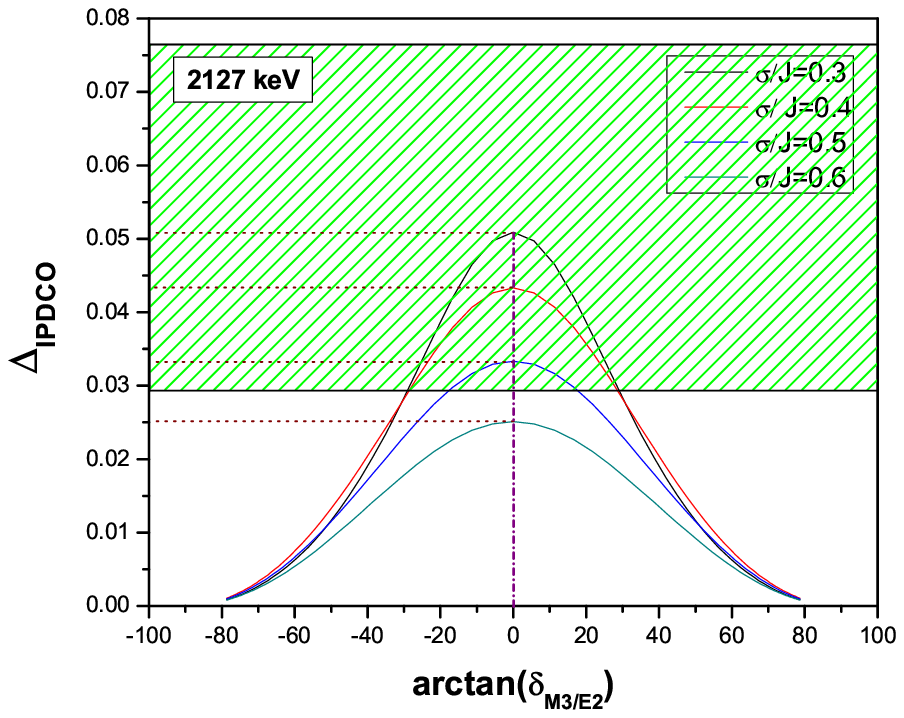}
\caption{\label{fig:8}``(Color online)" Plot of theoretical $\Delta_{IPDCO}$ as a function of mixing
ratios at different $\sigma$/J for 2127 keV (2$^{+}$ $\rightarrow$ 0$^{+}$)
in $^{34}$S. The shaded area represents the range of experimentally
measured $\Delta_{IPDCO}$. The shell model predicted mixing ratio
and the corresponding $\Delta_{IPDCO}$ values at different $\sigma$/J are marked by
the vertical and the horizontal dotted lines respectively.}
\end{figure}

Fig.~\ref{fig:8} shows the theoretical asymmetry values
for a pure E2 transition ($J^\pi$ = 2$^{+}$ $\rightarrow$ 0$^{+}$)
as a function of mixing ratio at different values of $\sigma/J$.
The shaded area gives the range of observed $\Delta_{IPDCO}$ values
for the known 2127-keV, E2 transition (2$^{+}$ $\rightarrow$ 0$^{+}$) in $^{34}$S.
An increase in $\sigma/J$ decreases polarization. At $\sigma/J$ = 0.6
the theoretical values are not consistent with the experimental values.
The theoretical asymmetry value at $\delta$ = 0 for $\sigma/J$ = 0.4 lies well within the 
experimentally observed range. This also justifies our choice of
$\sigma/J$.

The calculated and observed $R_{DCO}$ and $\Delta_{IPDCO}$ values for
several strong transitions in $^{34}$S, $^{33}$P and $^{33}$S are given in Table I.
The $R_{DCO}$ and $\Delta_{IPDCO}$ values are
consistently reproduced within error bars in each case.

Wherever quantitative measurements were not possible due to insufficient statistics,
parity assignment was done qualitatively from the corresponding
gated perpendicular and parallel spectra. When gates were put on coincident
gamma transitions and more counts were observed in the perpendicular
gated spectra than in the parallel, the observed $\gamma$-ray was
assigned an electric nature. The reverse was true for assignment
of a magnetic nature. 
Thus the previously reported
spin-parities of the levels in $^{33,34}$S and $^{33,34}$P that were observed in
the present experiment, have been confirmed either quantitatively or qualitatively.

\subsection{Level schemes} 

\begin{figure}[htp]
\includegraphics[trim=0.0cm 0.0cm 0.0cm 0.0cm,scale=0.50,angle=270]{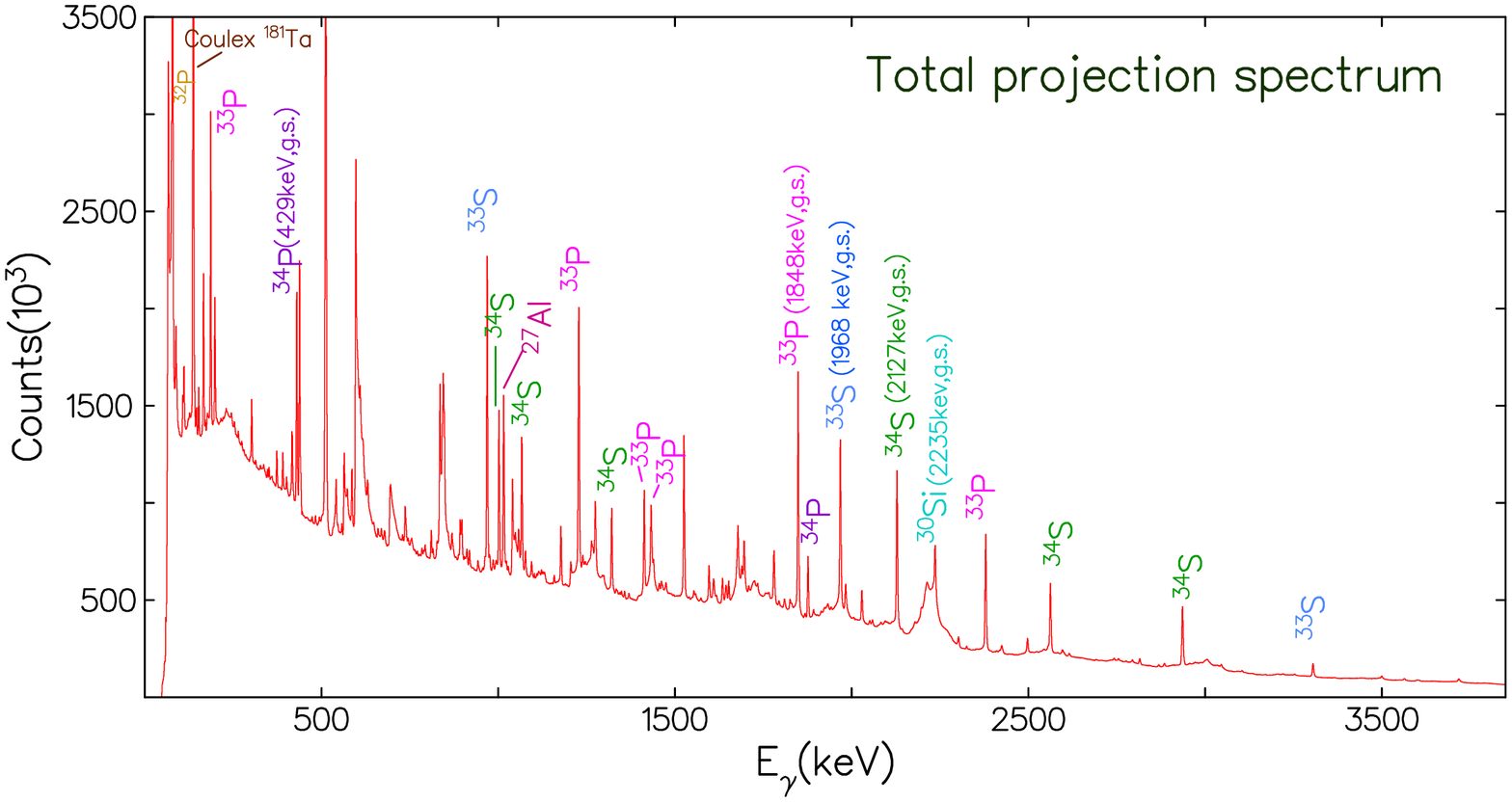}
\caption{\label{fig:9}Projection spectrum from $^{18}$O+$^{18}$O fusion
reaction at an incident beam energy of 34 MeV.}
\end{figure}

\begin{figure}[!ht]
\includegraphics[trim=0.0cm 0.0cm 0.0cm 0.0cm,scale=0.45,angle=270]{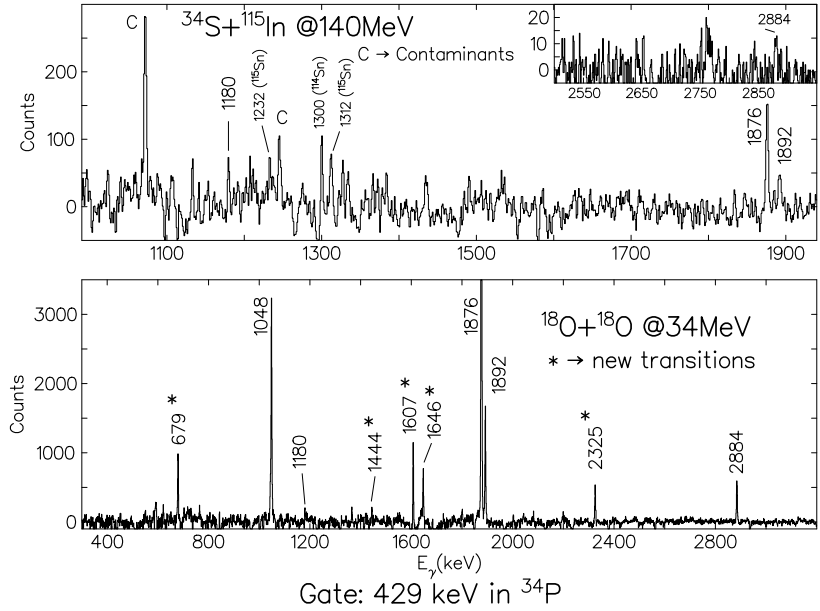}
\caption{\label{fig:10}Coincidence spectrum with gate set on 429 keV 
in $^{34}$P from: transfer/deep-inelastic reaction~\cite{Krishi} (top panel)
and present reaction (bottom panel). This figure highlights the advantage
of the fusion reaction over non-equilibrated reactions to populate nuclei
in and around the "island of inversion".}
\end{figure}

The nuclei populated in the experiment as
determined from the projection spectra
and supported by the beam-off radioactivity data
were $^{34}$S, $^{33}$S, $^{34}$P, $^{33}$P,
$^{32}$P and $^{30}$Si.
Fig.~\ref{fig:9} depicts the projection spectrum of the symmetric $\gamma$-$\gamma$
matrix. The use of fusion evaporation reaction to populate the above-mentioned nuclei
has clearly enhanced their production compared to
deep-inelastic/transfer reactions~\cite{Krishi} as is evident from Fig.~\ref{fig:10}.
The comparison also shows a much cleaner and contamination
free spectrum obtained in the present experiment.
Moreover, higher spin states have become accessible as
a result of utilizing fusion evaporation reaction mechanism.

\begin{figure}[htbp]
\includegraphics[trim=0.0cm 0.0cm 0.0cm 0.0cm, scale=0.5, angle=270]{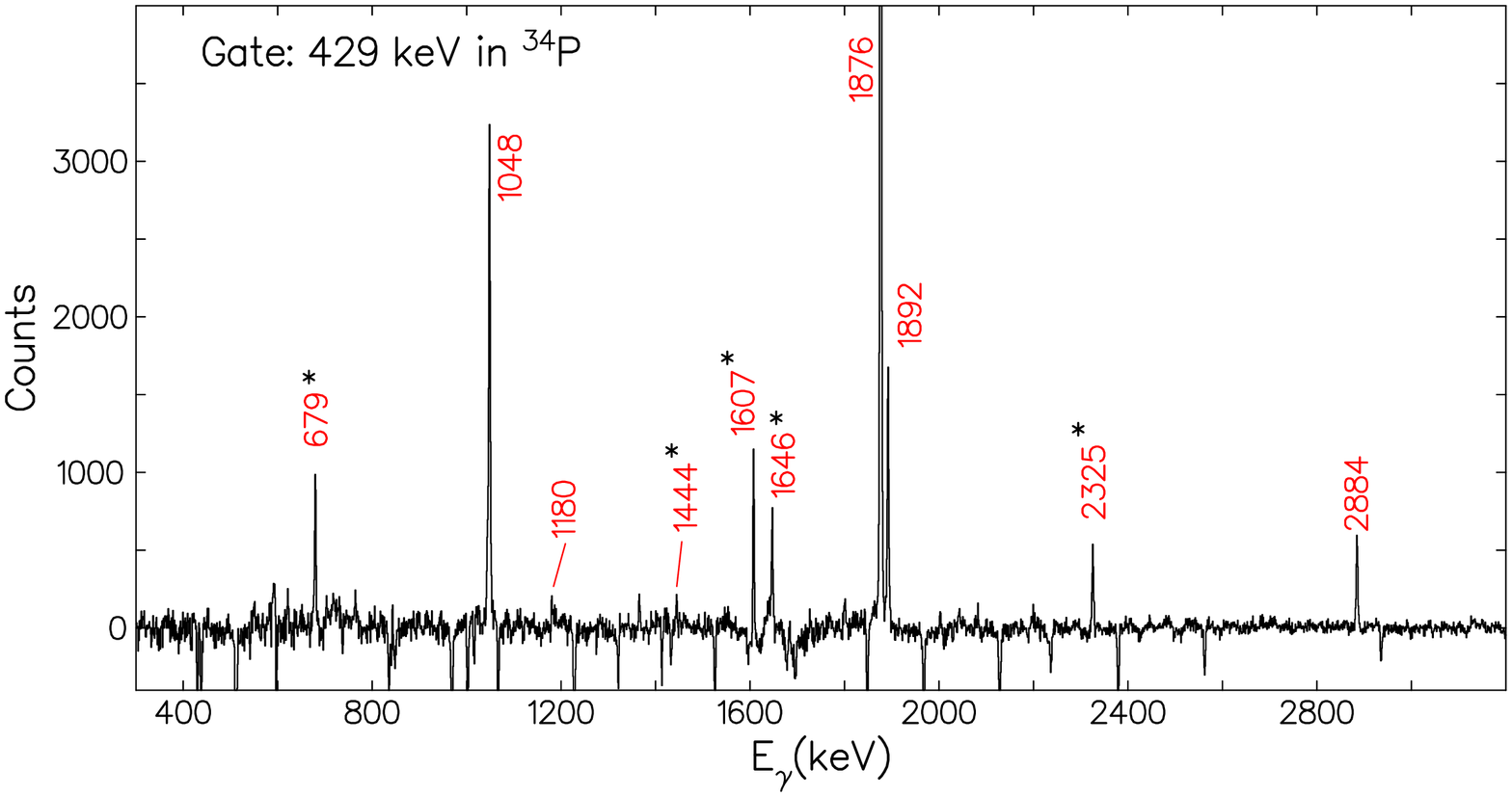}
\caption{\label{fig:11} ``(Color online)" Coincidence spectrum with gate on 429 keV in $^{34}$P.
The new assigned $\gamma$-rays are marked with an asterisk.}
\end{figure}

\begin{figure}[htbp]
\includegraphics[trim=0.0cm 0.0cm 0.0cm 0.0cm, scale=0.5, angle=270]{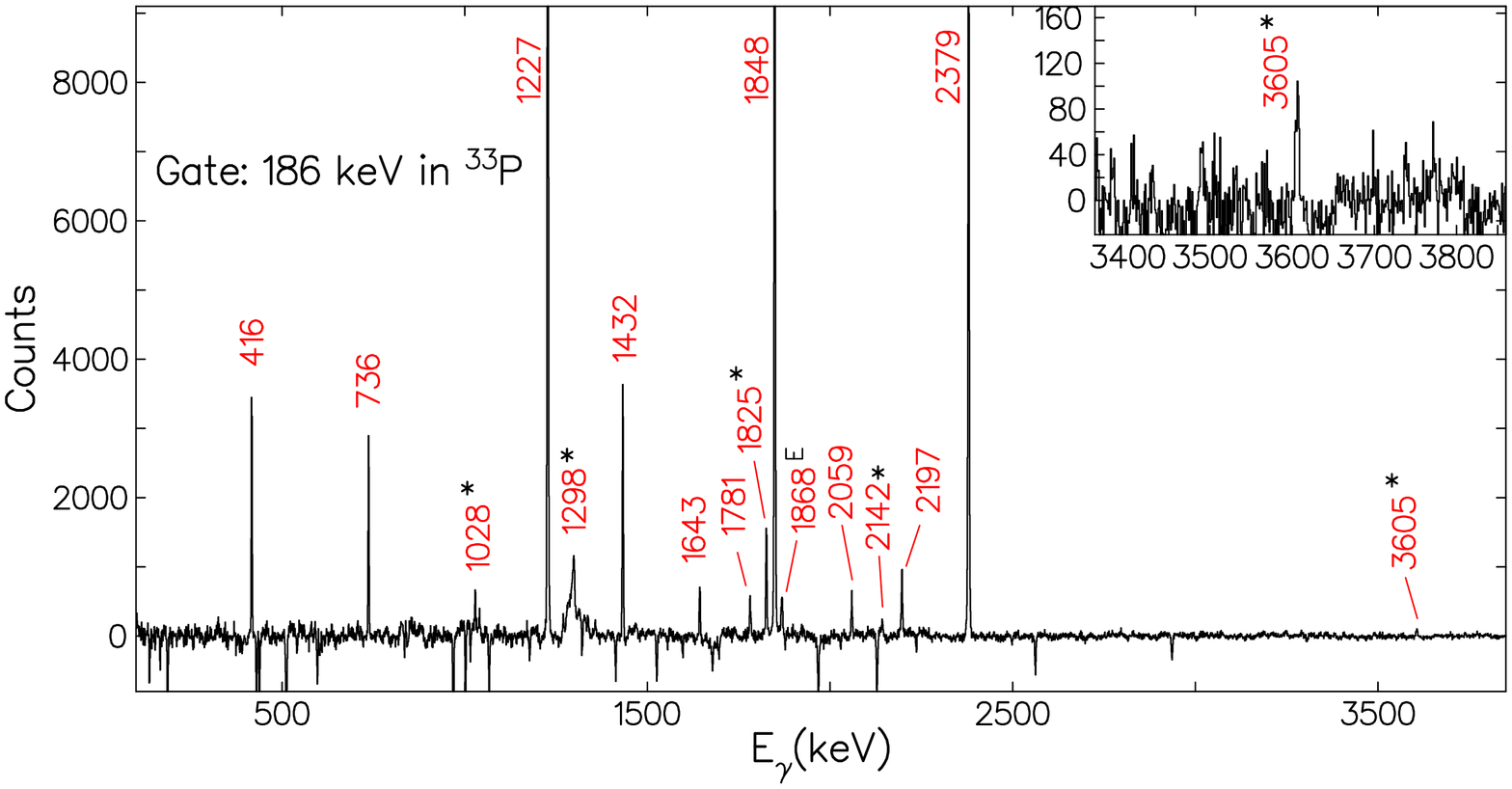}
\caption{\label{fig:12} ``(Color online)" Coincidence spectrum with gate on 186 keV in $^{33}$P.
The new assigned $\gamma$-rays are marked with an asterisk.}
\end{figure}

\begin{figure}[htbp]
\includegraphics[trim=0.0cm 0.0cm 0.0cm 0.0cm, scale=0.5,angle=270]{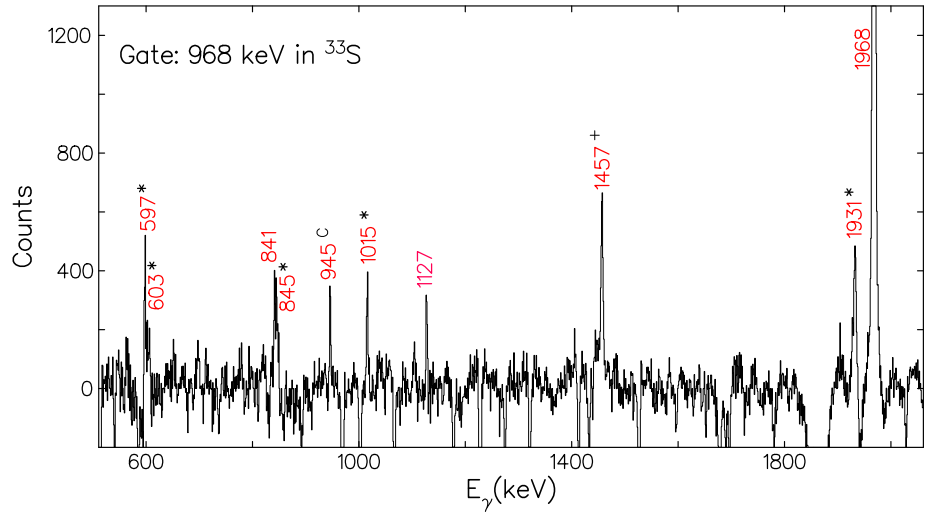}
\caption{\label{fig:13}``(Color online)" Coincidence spectrum with gate on 968 keV in $^{33}$S.
The new assigned $\gamma$-rays are marked with an asterisk. 945-keV is a contaminant
(C) from $^{30}$Si. 1457-keV(+) was found to be in coincidence with the 968-
and the 841-keV transitions, but could not be placed in the level scheme of $^{33}$S
from coincidence arguments.}
\end{figure}

\begin{figure}[htp]
\includegraphics[trim=3.0cm 2.0cm 0.0cm 0.0cm,scale=0.80,angle=0]{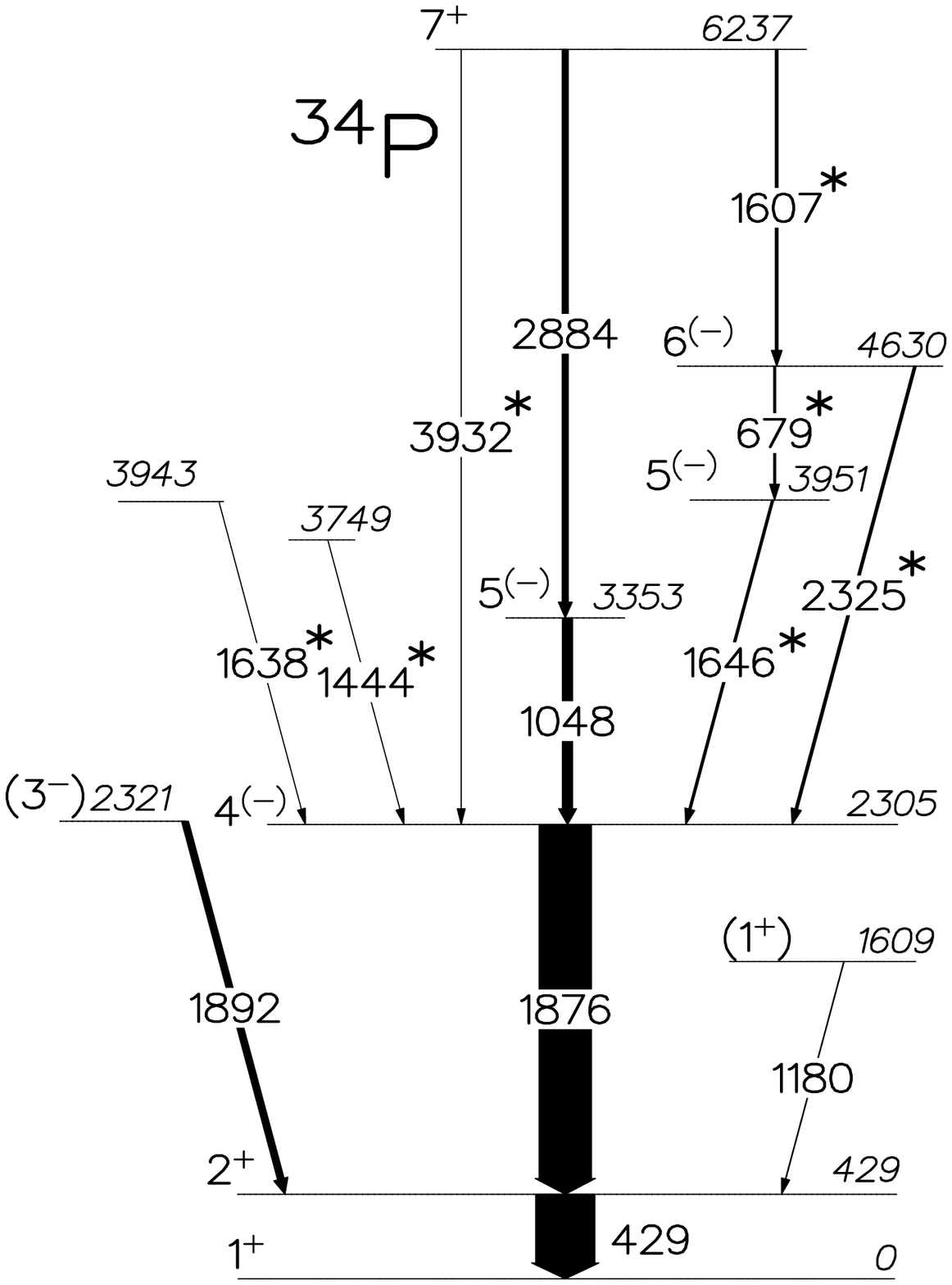}
\caption{\label{fig:14} Level scheme of $^{34}$P. The new transitions are 
indicated by an asterisk. The width of the arrows connecting the levels
is proportional to the relative $\gamma$-ray intensities.}
\end{figure}

\begin{figure}[htp]
\includegraphics[trim=0.0cm 0.0cm 0.0cm 0.0cm,scale=0.80,angle=0]{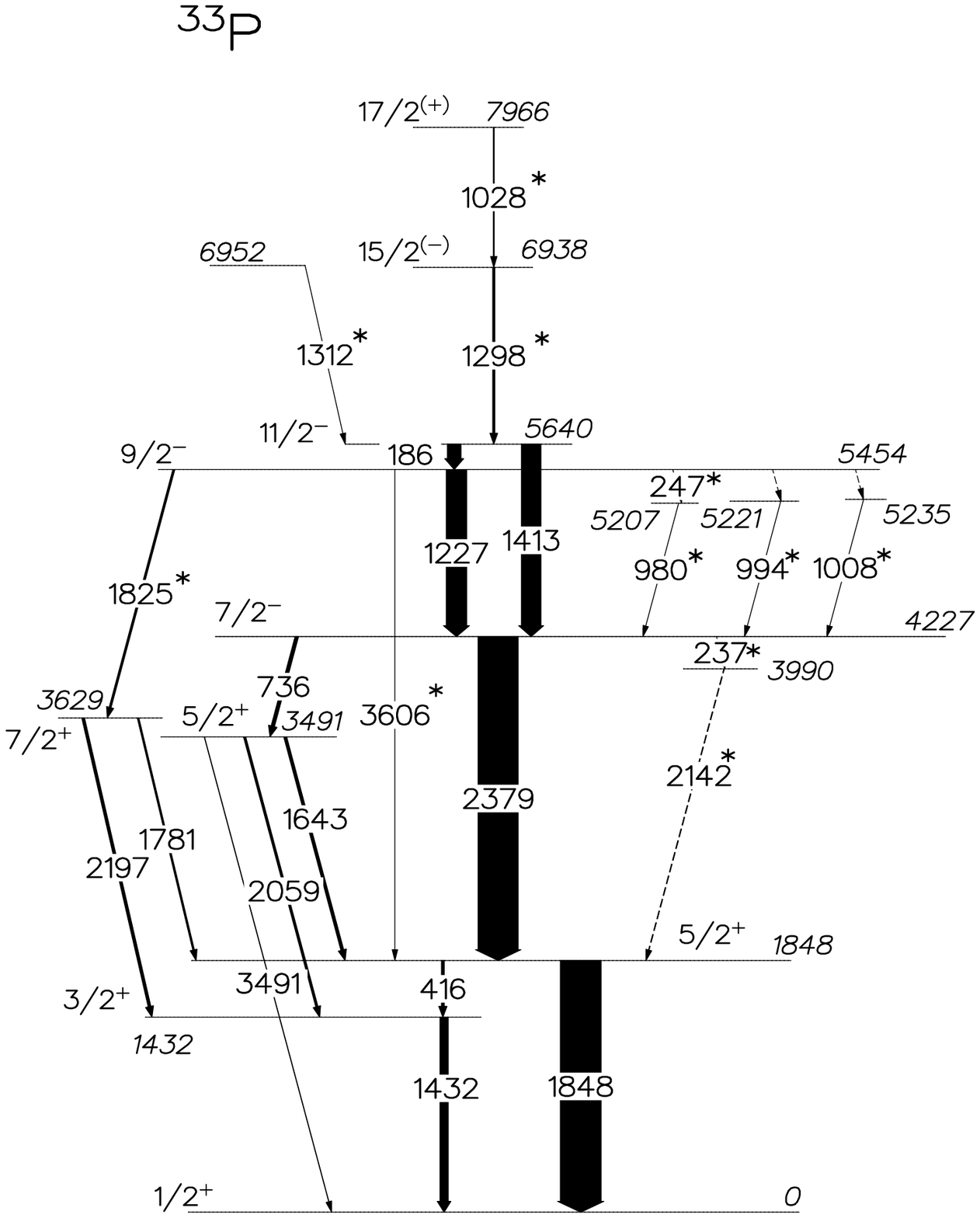}
\caption{\label{fig:15} Level scheme of $^{33}$P. The new transitions are
indicated by an asterisk. The width of the arrows connecting the levels
is proportional to the relative $\gamma$-ray intensities.}
\end{figure}

\begin{figure}[htp]
\includegraphics[trim=0.0cm 0.0cm 0.0cm 0.0cm,scale=0.80,angle=0]{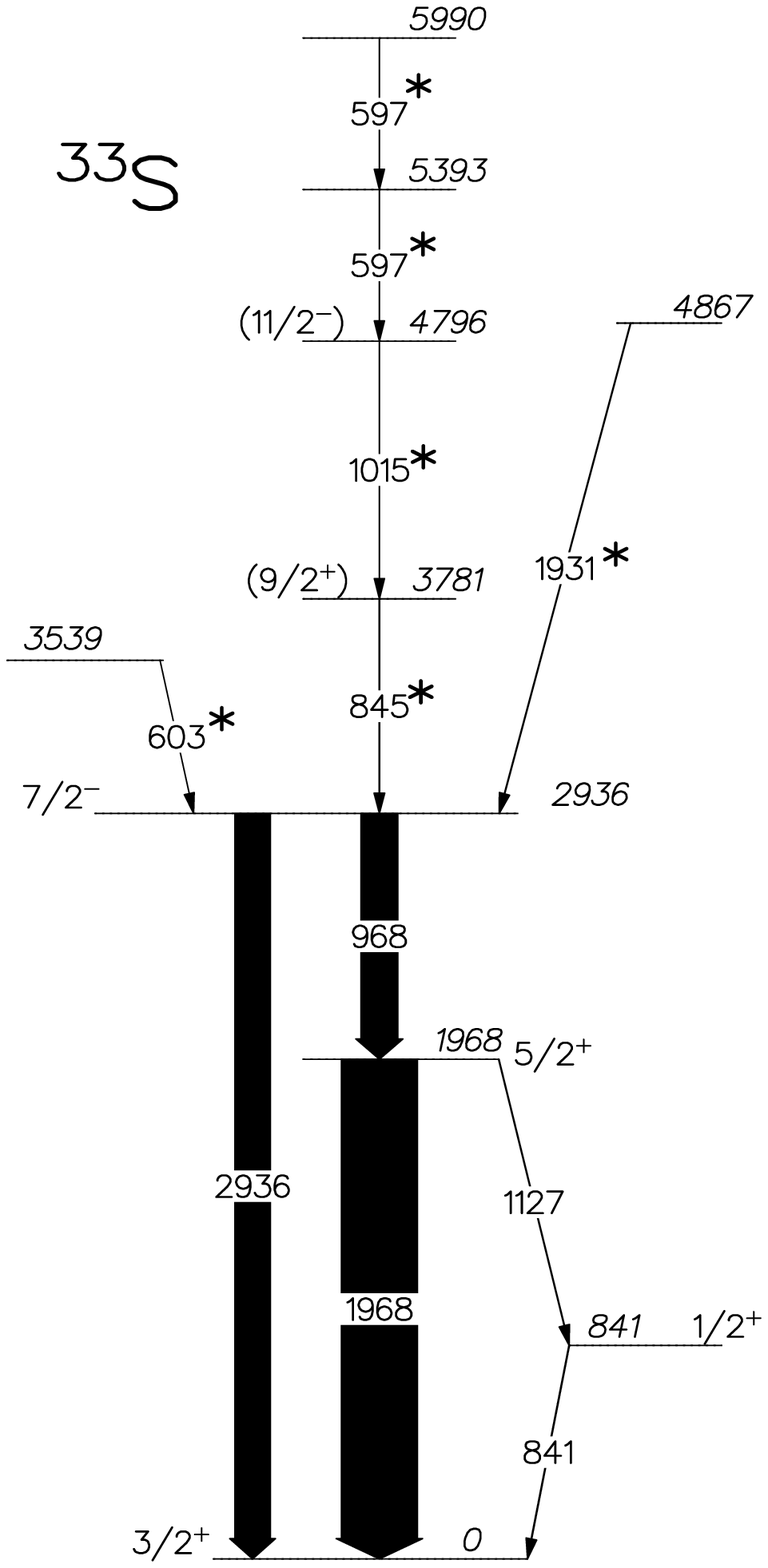}
\caption{\label{fig:16} Level scheme of $^{33}$S. The new transitions are
indicated by an asterisk. The width of the arrows connecting the levels
is proportional to the relative $\gamma$-ray intensities.}
\end{figure}

The level schemes of these nuclei have been extended with 
the addition of several new transitions. 
Figs.~\ref{fig:11},~\ref{fig:12} and~\ref{fig:13} are spectra 
obtained by gating on the symmetric $\gamma$-
$\gamma$ matrix by the 429-keV, 186-keV, and 968-keV $\gamma$-rays
belonging to $^{34}$P, $^{33}$P, and $^{33}$S, respectively.
The deduced level schemes are shown in Figs.~\ref{fig:14},
\ref{fig:15} and \ref{fig:16}. 
The energies, relative intensities and assigned
multipolarities of the observed transitions,
the assigned excitation energies and spin-parities of the levels, 
and the $\gamma$-ray branching ratios for decay of those levels
are listed in Table II.

A quantitative measure of $R_{DCO}$ and/or linear polarization
has not been possible for some transitions. In those cases, a 
qualitative assignment has been made as explained in the 
previous section.
The multipolarities listed in 
TABLE II for such transitions are essentially the dominant 
multipolarity; the extent of mixing
could not be determined. 
The present setup did not permit us to obtain the
mixing ratios experimentally. However mixing ratio range 
has been deduced from the $R_{DCO}$ and $\Delta_{IPDCO}$ values for 
some transitions in $^{34}$P, as explained later. 
The comparison between theoretical and experimental 
$R_{DCO}$ and $\Delta_{IPDCO}$ values in $^{34}$S, $^{33}$P, and $^{33}$S 
is based on previously-reported mixing ratios
(TABLE I).

\subsubsection{$^{34}$P}
Previous investigations of the level structure of $^{34}$P
employed non-equilibrated reactions to populate this difficult
to access N = 19 nucleus. These studies reported different
subsets of the excited levels in $^{34}$P.
Ajzenberg-Selove {\em et al.}~\cite{Ajzen} were the first to report the excited states of
$^{34}$P at 423, 1605, 2225, 2309 and (3345) keV($\pm$10 keV)
using the $^{34}$S(t,$^{3}$He)$^{34}$P reaction. On the other hand,
429-, 1608- and 1178-keV gamma rays were identified by both
Nathan {\em et al.}~\cite{Nathan} and Pritychenko {\em et al.}~\cite{Pritychenko}
in $\beta$-decay and intermediate-energy 
Coulomb excitation measurements respectively.
Pritychenko {\em et al.}~\cite{Pritychenko} also
observed a new 627-keV transition de-exciting
the (2$^{+}$) level at 2225 keV. However the 1608-keV
and 627-keV transition were not observed in any
of the subsequent investigations using transfer
and deep-inelastic reactions ~\cite{Fornal,Ollier,Krishi}.
All the excited states of $^{34}$P reported by
Ollier {\em et al.}~\cite{Ollier}, except the level
at 4723 keV were observed in the present work.
Several other strong transitions, of
679, 1444, 1607, 1638, 1646, 2325 and 3932 keV
belonging to $^{34}$P have been identified and placed
in the level scheme by coincidence and intensity arguments.
The placements of the new transitions were also
facilitated to a large extent by the observation of cross-over transitions.
The 1607-keV transition was found to be in coincidence with the 429- and 1876-keV
transitions and hence is different from the 1608-keV transition reported
by Nathan {\em et al.}~\cite{Nathan} and Pritychenko {\em et al.}
~\cite{Pritychenko}. The 1046-keV transition reported by
Ollier {\em et al.}~\cite{Ollier} but not 
observed by Krishichayan {\em et al.}~\cite{Krishi}
has been observed in the present experiment, but with an
energy of 1048-keV.  
Further, this transition exhibited a 
shape asymmetry at forward and backward angles, which is
indicative of a lifetime of the order of a few pico-second for the 3353 keV level.

\begin{figure}[htp]
\includegraphics[trim=0.0cm 0.0cm 0.0cm 0.0cm,scale=0.50,angle=270]{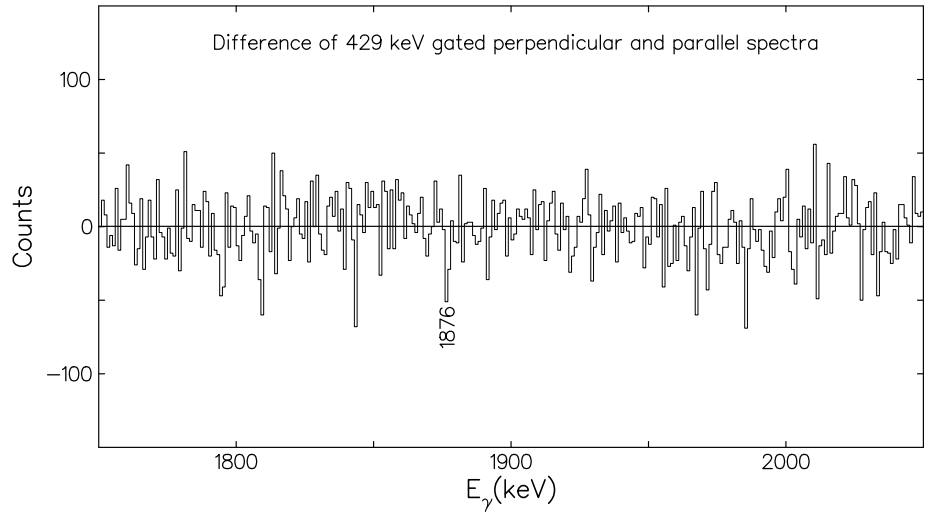}
\caption{\label{fig:17}Background subtracted difference spectrum of perpendicular
and parallel scattered events when gated by 429 keV transition in $^{34}P$. 
Absence of a clear positive or negative peak at 1876 keV is indicative of
its mixed nature.} 
\end{figure}

\begin{figure}[]
\includegraphics[trim=0.0cm 0.0cm 0.0cm 0.0cm, scale=0.9,angle=0]{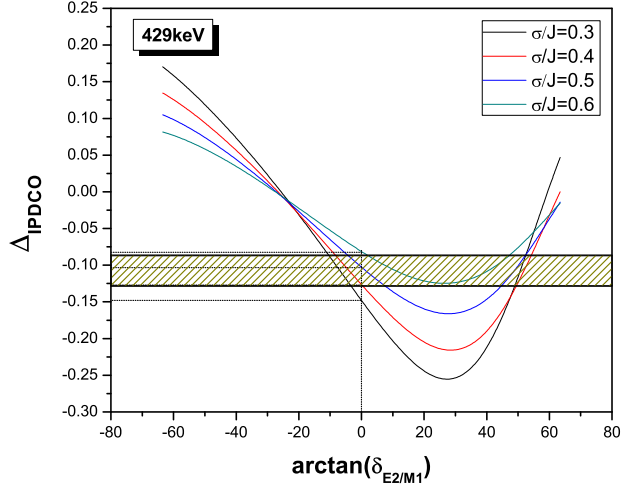}
\caption{\label{fig:18}``(Color online)" Plot of theoretical $\Delta_{IPDCO}$ as a function of mixing
ratios at different $\sigma/J$ for 429 keV (2$^{+}$ $\rightarrow$ 1$^{+}$) in $^{34}$P. The shaded 
area represents the range of experimentally measured $\Delta_{IPDCO}$. The shell model
predicted mixing ratio and the corresponding $\Delta_{IPDCO}$ values are marked by
the vertical and the horizontal dotted lines respectively.}
\end{figure}

\begin{figure}[]
\includegraphics[trim=0.0cm 0.0cm 0.0cm 0.0cm, scale=0.9,angle=0]{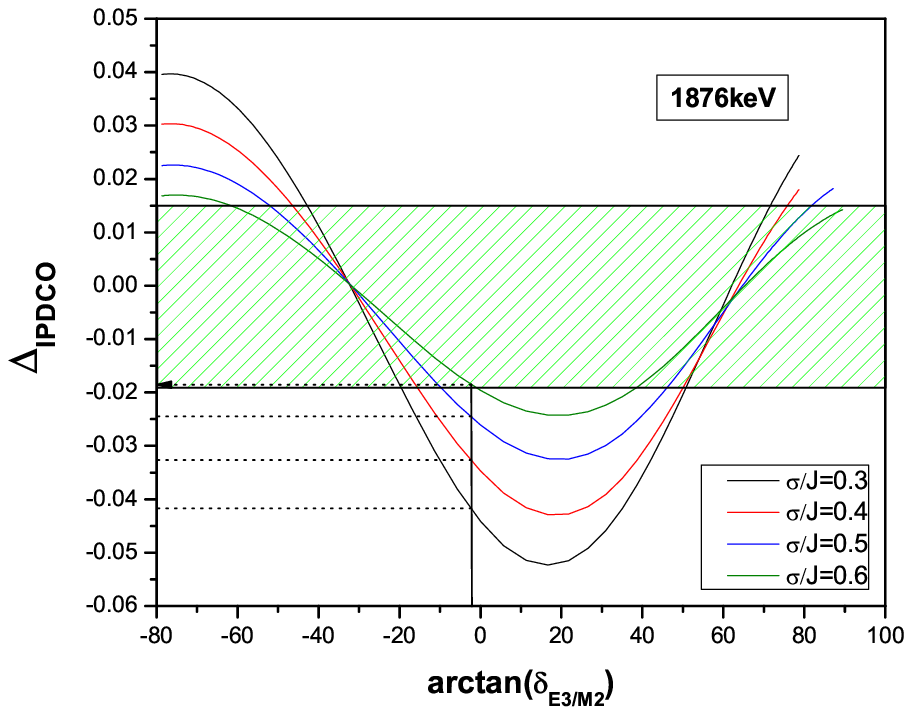}
\caption{\label{fig:19}``(Color online)" Plot of theoretical $\Delta_{IPDCO}$ as a function of mixing
ratios at different $\sigma/J$ for 1876 keV (4$^{-}$ $\rightarrow$ 2$^{+}$) in $^{34}$P considering
a M2+E3 distribution. The shaded area represents  
the range of experimentally measured $\Delta_{IPDCO}$. The shell model
predicted mixing ratio and the corresponding $\Delta_{IPDCO}$s values are marked by
the vertical and the horizontal dotted lines respectively.}
\end{figure}

One of the main motivations of the present experiment was
to undertake polarization and coincidence angular
correlation measurements following fusion reaction
to confirm the spin-parity assignment of the 2305-keV level
in $^{34}$P. In Ref.~\cite{Krishi},
we had assigned $J^\pi$ = 4$^{+}$ to this level,
whereas it had been assigned 4$^{-}$ by earlier workers
~\cite{Ajzen,Asai}. The 429-keV was established as a magnetic 
dipole transition from the DCO and polarization 
analysis method described in the previous section. 
The 1876-keV $\gamma$ ray de-exciting the 2305-keV
level has a $R_{DCO}$ value 1.62(26) (Fig.~\ref{fig:RDCOd})
and linear polarization measurements yielded a near-zero value
for $\Delta_{IPDCO}$, establishing it as a highly mixed
(L = 2, L$^{'}$ = 3) transition for the first time (Fig.~\ref{fig:19}). This is also evident from
the difference between the 429-keV gated perpendicular and
parallel scattering spectra (Fig.~\ref{fig:17}), where the number of counts under 1876 keV
peak is nearly zero. In such cases, the polarization measurements
cannot distinguish between M2/E3 and E2/M3 mixing.
Fig.~\ref{fig:18} depicts the theoretical asymmetry
curves for 2$^{+}$ $\rightarrow$ 1$^{+}$ M1/E2 radiation
as a function of mixing ratio, at different values of $\sigma/J$.
The shaded area represents the experimental dispersion in
$\Delta_{IPDCO}$ of the 429-keV $\gamma$ ray.
At $\sigma/J$ = 0.4, the theoretical values are consistent
with experimental $\Delta_{IPDCO}$ values
over the range -8$^{0} \leq$ arctan($\delta$)
$\leq$ 0$^{0}$ and 49$^{0} \leq$ arctan($\delta$) $\leq$ 54$^{0}$.
However, as seen from Fig.~\ref{fig:18}, the mixing ratio predicted by
the shell model (as detailed in the subsequent section
and indicated by the vertical dotted line) limits the value to
the former range.
We have similarly plotted the theoretical asymmetry values
for a M2/E3 mixing 4$^{-}$ $\rightarrow$ 2$^{+}$
as a function of mixing ratio, at different values of
$\sigma/J$ (Fig.~\ref{fig:19}). The shaded area on this graph represents
experimental range of $\Delta_{IPDCO}$ for the 1876-keV
transition. The theoretical values are consistent with the
experimental values over the range -46$^{0} \leq$ arctan($\delta$)
$\leq$ -15$^{0}$ and 50$^{0} \leq$ arctan($\delta$)
$\leq$ 76$^{0}$ for $\sigma/J$ = 0.4. The latter
range of very large mixing ratios being
physically unreasonable, has not been considered.
When we repeated this exercise considering a E2/M3
mixing, we obtained almost the same ranges of mixing
ratios. Thus an unambiguous identification
of the 1876-keV transition as M2/E3 or E2/M3 is not
possible following this procedure.
The $\Delta_{IPDCO}$ value  at $\sigma/J$ = 0.4
corresponding to the mixing ratio predicted by the shell model
calculations (Section IV)
does not lie within the aforementioned ranges
in both cases. This mismatch has been discussed in detail in
the subsequent section. 

\begin{figure}[]
\includegraphics[trim=0.0cm 0.0cm 0.0cm 0.0cm, scale=0.6,angle=270]{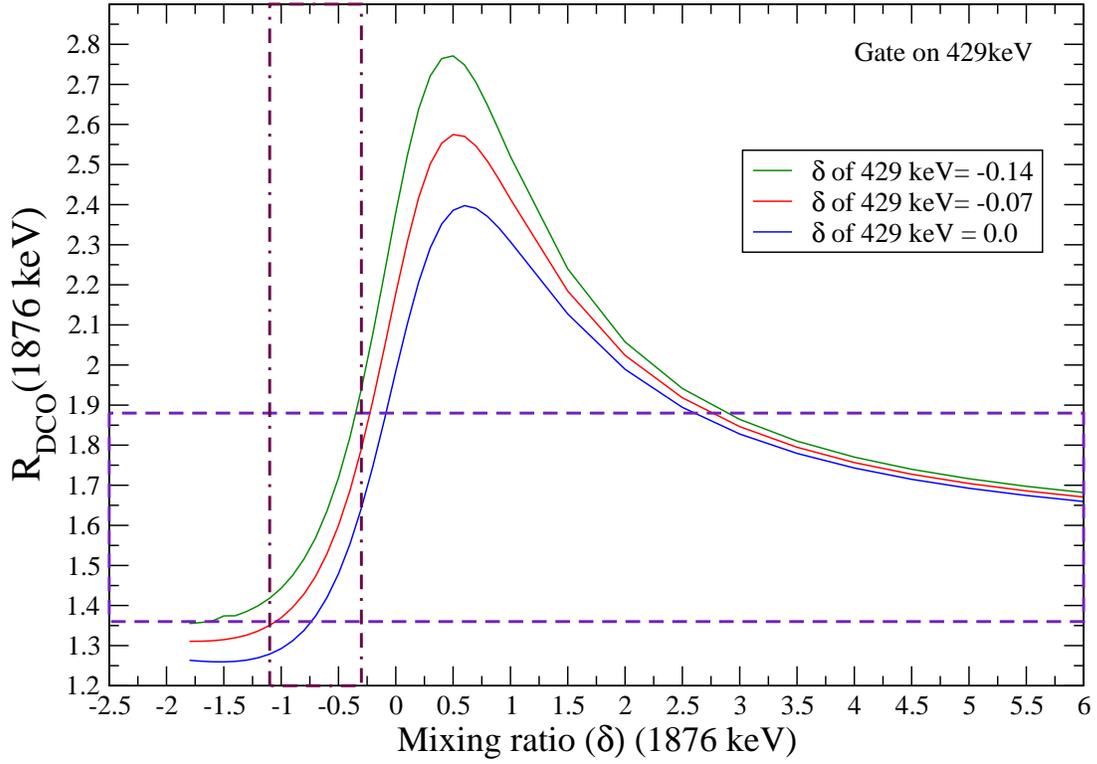}
\caption{\label{fig:20} ``(Color online)" The variation in $R_{DCO}$ for 1876 keV as a function
of its mixing ratio at
3 different values of mixing ratio of 429 keV (the gating transition).} 
\end{figure}

Fig.~\ref{fig:20} shows the variation of theoretical $R_{DCO}$
for the 1876-keV $\gamma$-ray as a function of its mixing ratio when the gate is on
the 429-keV, the ground state transition. The three plots correspond
to the three values of $\delta$$_{429}$ [ -0.14 $\leq$
-0.07 $\leq$ 0.0] that were determined earlier from Fig.~\ref{fig:18}.
The theoretical $R_{DCO}$ values were computed
using ANGCOR~\cite{Krane2}.
The horizontal lines mark the experimental range of $R_{DCO}$
values for the 1876-keV transition, while the vertical lines indicate the range of
mixing ratios for this transition as obtained from Fig.~\ref{fig:19}.
As is evident from the graph, the mixing ratio range
obtained from the analysis of the linear polarization measurements is also consistent
with the angular correlation measurements. Thus, both these
measurements are indicative of -1.03$\leq$ $\delta_{1876}$ $\leq$ -0.27.

Asai {\em et al.}~\cite{Asai} have reported the lifetime of the level
at $E_{x}$ = 2.305 MeV as 0.3 ns $\leq$ $t_{1/2}$ $\leq$ 2.5 ns.
Combining this lifetime measurement
with the mixing ratio range, that we have obtained for
1876-keV, we have calculated the experimental
reduced transition probabilities assuming both M2/E3
and E2/M3 mixing. The calculations are presented in TABLE III.
As is evident from the table, the lifetime measurements lead
to unacceptable M3 strengths~\cite{Skorka}. This supports an
M2/E3 assignment for the 1876-keV transition
and $J^\pi$ = 4$^{(-)}$
to the 2305 keV level. This needs to be confirmed with precise
lifetime measurements, however.

The qualitative linear polarization measurements for the
1607-, 1646-, 679- and 2325-keV transitions indicate an electric nature
for them. The spin-parity assignments for the
levels de-exciting via these transitions are based on the assumption
of $J^\pi$ = 4$^{(-)}$ for the 2305-keV level.

\subsubsection{$^{33}$P}

This work reports the first polarization measurement for
$^{33}$P populated in a heavy-ion fusion reaction.
The earlier light-ion induced reaction investigations
had established the level structure up to
a spin of J$^{\pi}$= 11/2$^{-}$ and E$_{x}$ $\sim$5.6 MeV
~\cite{Nixon}.
We have been able to extend the yrast sequence up to
J$^{\pi}$= 17/2$^{(+)}$ and E$_{x}$ $\sim$8 MeV 
due to the observation of two new transitions of energy 1298 keV (E2)
and 1028 keV (M1). The multipolarities of these two transitions
were assigned as quadrupole and dipole, respectively, on the 
basis of the observed $R_{DCO}$ values. 
The parity measurements have been shown as tentative since 
only qualitative measurements were possible. The 1028-keV
transition was identified as a magnetic transition
due to its preferential scattering in the parallel
direction as observed in 1298-keV gated perpendicular
and parallel spectra. On the other hand, the 1298-keV transition was assigned
an electric nature due to its preferential scattering in 
perpendicular direction. 
A full Doppler shift has been observed in the 1298-keV transition 
and hence, the lifetime of the 6938 keV 
level is expected to be much less as compared to the stopping time ($\sim$pico-second).
Apart from these, several other new transitions, with energies (in
ascending order) of 237, 247, 980, 994, 1008, 1312,
1825 (D), 2142 (Q) and 3605 keV were observed and placed 
in the decay scheme.
The present statistics did not permit us to observe the weak 
transition to the E$_{x}$ $\sim$5221-keV
from the level at 5454 keV. We have also observed 1170-
and 880-keV transitions which could not be placed in the decay 
scheme. It is worth mentioning that the single- (1868 keV)
and double-escape (1358 keV) peaks corresponding to the 
2379-keV transition were observed and the intensity of this
$\gamma$-ray reported in TABLE II 
was obtained from the sum of the counts under the full photopeak and 
the two escape peaks. This was done as our efficiency  measurements
(performed with a $^{152}$Eu source) did not have data points in this 
energy region where escape
contribution becomes significant.

\subsubsection{$^{33}$S}

Prior to this experiment $^{33}$S has been studied 
via light ion reactions~\cite{Sterrenburg}.
In this experiment, the level scheme of $^{33}$S was 
extended with the addition of 6 
new transitions of energy 597, 597, 603, 845 (E1), 1015 (E1) and 
1931 keV respectively. 
The presence of a 597-keV transition in the 597-keV gated spectrum
is indicative of a doublet. 
As a result it was not possible to determine
their individual intensities and the spin parity of the excited 
states that de-excite via these two transitions. The qualitative
polarization and angular correlation measurements for the 845-keV and the 1015-keV
transitions indicate that these are dipoles and electric in nature.
Such sequences of electric transitions have been reported in neighbouring
nuclei like $^{32}$P~\cite{NNDC}.
The 1931-keV transition exhibits the fully Doppler shifted peak indicating a short
lifetime ($\ll$ pico-second, the stopping time) for the level at $\sim$4867 keV.
The present polarization and angular correlation
measurements confirmed the previously assigned spin-parity of the 
1968- and 2936-keV levels.
The reported mixing ratios are consistent
with the results of the present measurement (TABLE I). 

\section{THEORETICAL RESULTS}

Shell model calculations using the code NuShell@MSU~\cite{NuShell}
were performed to interpret the observed level structures 
of $^{33,34}$P and $^{33}$S. The valence space consisted of the
$1d_{5/2}$, $1d_{3/2}$, $2s_{1/2}$, $1f_{7/2}$, $1f_{5/2}$, $2p_{3/2}$ and
$2p_{1/2}$ orbitals outside a $^{16}$O core.
The ``{\em sdpfmw}'' interaction, taken from the Warburton, Becker,
Millener, and Brown (WBMB) {\em sd-pf} shell Hamiltonian~\cite{Warburton}, was used.

In $^{34}$P the positive-parity states 1$^{+}$ and 2$^{+}$, which
are expected to be dominated by the pure {\em sd} configurations, are well 
reproduced within the full {\em sd}-space shell model
calculations (0$\hbar$$\omega$) (using the {\em sdpfmw} interaction) and 
are consistent with the {\em sd} calculation of Brown~\cite{Brown}.
The predicted binding energy of the ground state is -191.971 MeV,
which matches very well with the experimental value -192.04 MeV~\cite{Brown}.
The mixing ratio of the 429-keV transition predicted by shell model
is -0.0024, which is also within the range determined
from our polarization measurements. 

Excitations of nucleons from {\em sd} shell into {\em fp} shell are essential to
explain the negative-parity states (minimum 1 particle in the {\em fp} shell)
as well as the high-spin, positive-parity states
(minimum 2 particles in the {\em fp} shell).
Due to computational limitations, unrestricted calculations 
were not possible and only one particle could be excited to the
pf shell (1$\hbar$$\omega$).   

It has been reported by several
authors that there is an overestimation of the {\em sd-pf} gap
in the corresponding interaction which required the lowering
of the single-particle energies of the {\em f} and {\em p} orbitals~\cite{Bujor,Mason}.
No such attempt was made in the present calculation.

\begin{figure}[htbp]
\includegraphics[trim=0.0cm 0.0cm 0.0cm 0.0cm, scale=0.8,angle=0]{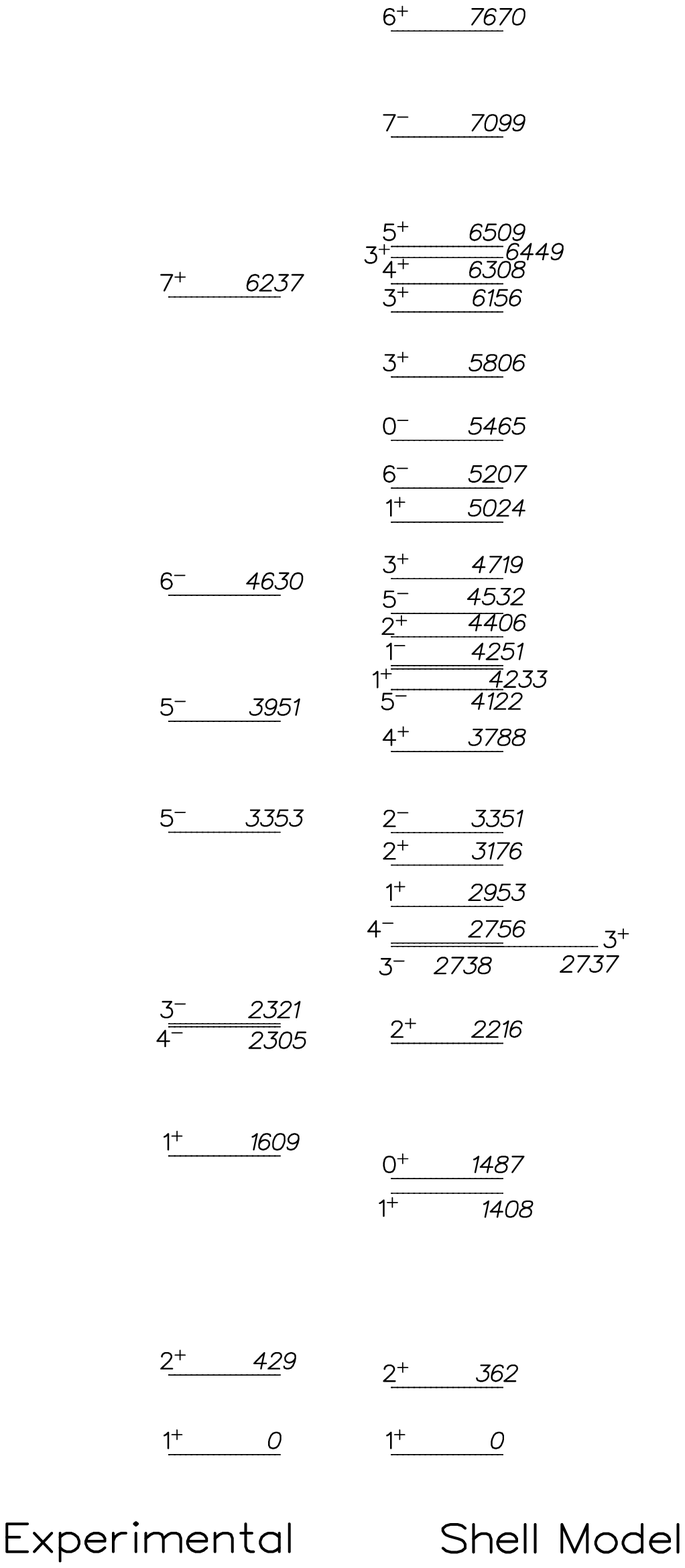}
\caption{\label{fig:sm} Comparison between experimental and
shell model predicted levels in $^{34}$P.}
\end{figure}

Fig.~\ref{fig:sm} shows a the comparison between the calculated and
the experimental levels in $^{34}$P. The 7$^{+}$ state predicted by the shell
model is at a very high excitation energy (11366 keV) and hence 
has not been included in the figure.
As seen from Fig.~\ref{fig:sm}, the high-spin positive-parity
states are much higher in excitation energy than the corresponding experimental
levels. This is likely due to our inability to excite more
than one particle into the {\em fp} shell.
There is a reasonable agreement in excitation energy
between the $J^\pi$ = 4$^{-}$, 5$^{-}$, 6$^{-}$ levels predicted by shell model 
and the observed 2305-, 3353- and 4630- keV levels, respectively.
Thus, the theory corroborates our spin-parity assignments at 
least for the negative-parity states.
However, the above shell model calculations failed to predict the mixed
nature of the 1876-keV transition established from our polarization measurements.
The shell model predicts an almost pure M2 nature for this transition
[$\delta$ = -0.034].
The B(M2) and B(E3) values obtained from shell model are 
0.1816 W.u. and 0.2167 W.u., respectively, and, as seen from
Table II, the B(E3) values are heavily under predicted,
clearly reflecting this mismatch.
We have also performed similar shell model calculations
for the neighbouring N = 19 isotones viz., $^{37}$Ar and $^{35}$S
where similar M2/E3 mixed transitions are reported (1611 keV ($J^\pi$ = 
7/2$^{-}$ $\rightarrow$ 3/2$^{-}$ in $^{37}$Ar) and 1911 keV ($J^\pi$ = 7/2$^{-}$
$\rightarrow$ 3/2$^{-}$ in $^{35}$S)~\cite{NNDC}. The results are summarized in TABLE IV. 
In all cases, the calculations  predict 
very little mixing, unlike the experimental observations.
The E3 transition strengths are several orders of magnitude higher
than the corresponding shell model predictions.
Clearly, there is a need to perform 
these calculations with a better Hamiltonian encompassing
a realistic cross-shell interaction, and/or with a more
complete wave function incorporating configurations arising
from multi-particle excitations into the fp orbitals.

The 0$\hbar$$\omega$ calculations for $^{33}$P and $^{33}$S reproduces
the low-spin positive-parity states. However, the 1$\hbar$$\omega$  
calculations fail to generate the first experimentally-observed negative-
parity state, 7/2$^{-}$, in both nuclei. The predicted energies of 
the high-spin, negative-parity states are higher than the 
experimental values by several MeV.

\section{CONCLUSIONS}
The level structure of the generally difficult to access nuclei
$^{33,34}$P and $^{33}$S has been investigated using heavy-ion
fusion reaction which has resulted in a substantial enhancement
in their production cross-sections. The level schemes of these nuclei
have been considerably extended. Spin-parity 
assignments have been made following a consistent analysis of
both the coincidence angular correlation and linear polarization
data. The results indicate that the 1876-keV transition
de-exciting the 2305-keV level in $^{34}$P is a mixed
transition and plausibly has a M2/E3 admixture; however
precise lifetime measurements would be required to confirm this
assignment unambiguously. 
The shell model calculations emphasize the need for detailed 
microscopic calculations to understand the observed level sequences and
mixing ratios. The deformed shell model could provide an insight 
into the observed level structures due to the 
occupation of deformation-driving orbitals such as $f_{7/2}$.

\section{ACKNOWLEDGEMENTS}
The authors would like to thank all the participants who have helped
set up the Clover array at TIFR. The help and co-operation received
from Mr Kaushik Basu of UGC-DAE CSR during the experiment is gratefully acknowledged.  
We would like to thank the BARC-TIFR Pelletron staff for their excellent
support during the experiment. We are thankful to  
Mr. J. P. Greene, ANL, U.S.A, for the $^{18}$O target. Thanks are
also due to Dr. W. P. Tan and Dr. Larry Lamm, Univ. of
Notre Dame, U.S.A, for providing us the enriched $^{18}$O cathode.
Special thanks to Prof. Alex Brown for the indepth 
discussions and his views and
comments on Shell model calculations.

\bibliography{nrich}

\begin{thebibliography}{37}
\expandafter\ifx\csname natexlab\endcsname\relax\def\natexlab#1{#1}\fi
\expandafter\ifx\csname bibnamefont\endcsname\relax
  \def\bibnamefont#1{#1}\fi
\expandafter\ifx\csname bibfnamefont\endcsname\relax
  \def\bibfnamefont#1{#1}\fi
\expandafter\ifx\csname citenamefont\endcsname\relax
  \def\citenamefont#1{#1}\fi
\expandafter\ifx\csname url\endcsname\relax
  \def\url#1{\texttt{#1}}\fi
\expandafter\ifx\csname urlprefix\endcsname\relax\def\urlprefix{URL }\fi
\providecommand{\bibinfo}[2]{#2}
\providecommand{\eprint}[2][]{\url{#2}}

\bibitem[{\citenamefont{{Fornal {\em et al.}}}(1994)}]{Fornal}
\bibinfo{author}{\bibfnamefont{B.}~\bibnamefont{{Fornal {\em et al.}}}},
  \bibinfo{journal}{Phys.\ Rev.\ C} \textbf{\bibinfo{volume}{49}},
  \bibinfo{pages}{2413} (\bibinfo{year}{1994}).

\bibitem[{\citenamefont{Asai et~al.}()\citenamefont{Asai, Ishii, Makishima,
  Ogawa, and Matsuda}}]{Asai}
\bibinfo{author}{\bibfnamefont{M.}~\bibnamefont{Asai}},
  \bibinfo{author}{\bibfnamefont{T.}~\bibnamefont{Ishii}},
  \bibinfo{author}{\bibfnamefont{A.}~\bibnamefont{Makishima}},
  \bibinfo{author}{\bibfnamefont{M.}~\bibnamefont{Ogawa}}, \bibnamefont{and}
  \bibinfo{author}{\bibfnamefont{M.}~\bibnamefont{Matsuda}},
  \eprint{Proceedings of the Third International Conference on Fission and
  Properties of Neutron-Rich Nuclei, edited by J.~H.~Hamilton, A.~V.~Ramayya,
  H.~K.~Carter (World Scientific, Singapore, 2002)~pp.~295-297}.

\bibitem[{\citenamefont{{Ollier {\em et al.}}}(2005)}]{Ollier}
\bibinfo{author}{\bibfnamefont{J.}~\bibnamefont{{Ollier {\em et al.}}}},
  \bibinfo{journal}{Phys.\ Rev.\ C} \textbf{\bibinfo{volume}{71}},
  \bibinfo{pages}{034316} (\bibinfo{year}{2005}).

\bibitem[{\citenamefont{Broda}(2006)}]{Broda}
\bibinfo{author}{\bibfnamefont{R.}~\bibnamefont{Broda}}, \bibinfo{journal}{J.\
  Phys.\ G:\ Nucl.\ Part.\ Phys.} \textbf{\bibinfo{volume}{32}},
  \bibinfo{pages}{R151} (\bibinfo{year}{2006}).

\bibitem[{\citenamefont{{Krishichayan {\em et al.}}}(2006)}]{Krishi}
\bibinfo{author}{\bibnamefont{{Krishichayan {\em et al.}}}},
  \bibinfo{journal}{Eur. Phys. J. A} \textbf{\bibinfo{volume}{29}},
  \bibinfo{pages}{151} (\bibinfo{year}{2006}).

\bibitem[{\citenamefont{Nathan and Alburger}(1977)}]{Nathan}
\bibinfo{author}{\bibfnamefont{A.~M.} \bibnamefont{Nathan}} \bibnamefont{and}
  \bibinfo{author}{\bibfnamefont{D.~E.} \bibnamefont{Alburger}},
  \bibinfo{journal}{Phys.\ Rev.\ C} \textbf{\bibinfo{volume}{15}},
  \bibinfo{pages}{1448} (\bibinfo{year}{1977}).

\bibitem[{\citenamefont{{Iwasa {\em et al.}}}(2003)}]{Iwasa}
\bibinfo{author}{\bibfnamefont{I.}~\bibnamefont{{Iwasa {\em et al.}}}},
  \bibinfo{journal}{Phys.\ Rev.\ C} \textbf{\bibinfo{volume}{67}},
  \bibinfo{pages}{064315} (\bibinfo{year}{2003}).

\bibitem[{\citenamefont{Eyal and Dastrovosky}(1972)}]{Eyal}
\bibinfo{author}{\bibfnamefont{Y.}~\bibnamefont{Eyal}} \bibnamefont{and}
  \bibinfo{author}{\bibfnamefont{I.}~\bibnamefont{Dastrovosky}},
  \bibinfo{journal}{Nucl.\ Phys.\ A} \textbf{\bibinfo{volume}{179}},
  \bibinfo{pages}{594} (\bibinfo{year}{1972}).

\bibitem[{LAM()}]{LAMPS}
\emph{\bibinfo{title}{{LAMPS}}},
  \bibinfo{note}{\url{http://www.tifr.res.in/~pell/lamps.html#}}.

\bibitem[{\citenamefont{Pattabiraman
  et~al.}(2004{\natexlab{a}})\citenamefont{Pattabiraman, Chintalapudi, and
  Ghugre}}]{nsp1}
\bibinfo{author}{\bibfnamefont{N.~S.} \bibnamefont{Pattabiraman}},
  \bibinfo{author}{\bibfnamefont{S.~N.} \bibnamefont{Chintalapudi}},
  \bibnamefont{and} \bibinfo{author}{\bibfnamefont{S.~S.}
  \bibnamefont{Ghugre}}, \bibinfo{journal}{Nucl.\ Instrum.\ Methods\ Phys.\
  Res.\ A,} \textbf{\bibinfo{volume}{526}}, \bibinfo{pages}{432}
  (\bibinfo{year}{2004}{\natexlab{a}}).

\bibitem[{\citenamefont{Pattabiraman
  et~al.}(2004{\natexlab{b}})\citenamefont{Pattabiraman, Chintalapudi, and
  Ghugre}}]{nsp2}
\bibinfo{author}{\bibfnamefont{N.~S.} \bibnamefont{Pattabiraman}},
  \bibinfo{author}{\bibfnamefont{S.~N.} \bibnamefont{Chintalapudi}},
  \bibnamefont{and} \bibinfo{author}{\bibfnamefont{S.~S.}
  \bibnamefont{Ghugre}}, \bibinfo{journal}{Nucl.\ Instrum.\ Methods\ Phys.\
  Res.\ A,} \textbf{\bibinfo{volume}{526}}, \bibinfo{pages}{439}
  (\bibinfo{year}{2004}{\natexlab{b}}).

\bibitem[{\citenamefont{Pattabiraman et~al.}(2006)\citenamefont{Pattabiraman,
  Ghugre, Basu, Garg, Ray, Sinha, and Zhu}}]{nsp3}
\bibinfo{author}{\bibfnamefont{N.~S.} \bibnamefont{Pattabiraman}},
  \bibinfo{author}{\bibfnamefont{S.~S.} \bibnamefont{Ghugre}},
  \bibinfo{author}{\bibfnamefont{S.~K.} \bibnamefont{Basu}},
  \bibinfo{author}{\bibfnamefont{U.}~\bibnamefont{Garg}},
  \bibinfo{author}{\bibfnamefont{S.}~\bibnamefont{Ray}},
  \bibinfo{author}{\bibfnamefont{A.~K.} \bibnamefont{Sinha}}, \bibnamefont{and}
  \bibinfo{author}{\bibfnamefont{S.}~\bibnamefont{Zhu}},
  \bibinfo{journal}{Nucl.\ Instrum.\ Methods\ Phys.\ Res.\ A,}
  \textbf{\bibinfo{volume}{562}}, \bibinfo{pages}{222} (\bibinfo{year}{2006}).

\bibitem[{\citenamefont{{C.~Radford}}(1995)}]{radford}
\bibinfo{author}{\bibfnamefont{D.}~\bibnamefont{{C.~Radford}}},
  \bibinfo{journal}{Nucl.\ Instrum.\ Methods\ Phys.\ Res.\ A,}
  \textbf{\bibinfo{volume}{361}}, \bibinfo{pages}{297} (\bibinfo{year}{1995}).

\bibitem[{\citenamefont{Krane and Steffen}(1970)}]{Krane}
\bibinfo{author}{\bibfnamefont{K.~S.} \bibnamefont{Krane}} \bibnamefont{and}
  \bibinfo{author}{\bibfnamefont{R.~M.} \bibnamefont{Steffen}},
  \bibinfo{journal}{Phys.\ Rev.\ C} \textbf{\bibinfo{volume}{2}},
  \bibinfo{pages}{724} (\bibinfo{year}{1970}).

\bibitem[{\citenamefont{{R. M.~Steffan} and {K.~Alder}}(1975)}]{Steffan}
\bibinfo{author}{\bibnamefont{{R. M.~Steffan}}} \bibnamefont{and}
  \bibinfo{author}{\bibnamefont{{K.~Alder}}}, \emph{\bibinfo{title}{The
  Electromagnetic Interaction in Nuclear Spectroscopy}}
  (\bibinfo{publisher}{North-Holland, Amsterdam}, \bibinfo{year}{1975}).

\bibitem[{\citenamefont{Fagg and Hanna}(1959)}]{Fagg}
\bibinfo{author}{\bibfnamefont{L.~W.} \bibnamefont{Fagg}} \bibnamefont{and}
  \bibinfo{author}{\bibfnamefont{S.~S.} \bibnamefont{Hanna}},
  \bibinfo{journal}{Rev.\ Mod.\ Phys.} \textbf{\bibinfo{volume}{31}},
  \bibinfo{pages}{711} (\bibinfo{year}{1959}).

\bibitem[{\citenamefont{Deng et~al.}(1992)\citenamefont{Deng, Ma, Hamilton,
  Ramayya, Rikovska, Stone, Croft, Piercey, Morgan, and {Mantica, Jr.}}}]{Deng}
\bibinfo{author}{\bibfnamefont{J.~K.} \bibnamefont{Deng}},
  \bibinfo{author}{\bibfnamefont{W.~C.} \bibnamefont{Ma}},
  \bibinfo{author}{\bibfnamefont{J.~H.} \bibnamefont{Hamilton}},
  \bibinfo{author}{\bibfnamefont{A.~V.} \bibnamefont{Ramayya}},
  \bibinfo{author}{\bibfnamefont{J.}~\bibnamefont{Rikovska}},
  \bibinfo{author}{\bibfnamefont{N.~J.} \bibnamefont{Stone}},
  \bibinfo{author}{\bibfnamefont{W.~L.} \bibnamefont{Croft}},
  \bibinfo{author}{\bibfnamefont{R.~B.} \bibnamefont{Piercey}},
  \bibinfo{author}{\bibfnamefont{J.~C.} \bibnamefont{Morgan}},
  \bibnamefont{and} \bibinfo{author}{\bibfnamefont{P.~F.}
  \bibnamefont{{Mantica, Jr.}}}, \bibinfo{journal}{Nucl.\ Instrum.\ Methods\
  Phys.\ Res.\ A,} \textbf{\bibinfo{volume}{317}}, \bibinfo{pages}{242}
  (\bibinfo{year}{1992}).

\bibitem[{\citenamefont{Jones et~al.}(1995)\citenamefont{Jones, Wei, Beck,
  Butler, Byrski, Duch\^ene, de~France, Hannachi, Jones, and Kharraja}}]{Jones}
\bibinfo{author}{\bibfnamefont{P.~M.} \bibnamefont{Jones}},
  \bibinfo{author}{\bibfnamefont{L.}~\bibnamefont{Wei}},
  \bibinfo{author}{\bibfnamefont{F.~A.} \bibnamefont{Beck}},
  \bibinfo{author}{\bibfnamefont{P.~A.} \bibnamefont{Butler}},
  \bibinfo{author}{\bibfnamefont{T.}~\bibnamefont{Byrski}},
  \bibinfo{author}{\bibfnamefont{G.}~\bibnamefont{Duch\^ene}},
  \bibinfo{author}{\bibfnamefont{G.}~\bibnamefont{de~France}},
  \bibinfo{author}{\bibfnamefont{F.}~\bibnamefont{Hannachi}},
  \bibinfo{author}{\bibfnamefont{G.~D.} \bibnamefont{Jones}}, \bibnamefont{and}
  \bibinfo{author}{\bibfnamefont{B.}~\bibnamefont{Kharraja}},
  \bibinfo{journal}{Nucl.\ Instrum.\ Methods\ Phys.\ Res.\ A,}
  \textbf{\bibinfo{volume}{362}}, \bibinfo{pages}{556} (\bibinfo{year}{1995}).

\bibitem[{\citenamefont{{Starosta {\em et al.}}}(1999)}]{Starosta}
\bibinfo{author}{\bibfnamefont{K.}~\bibnamefont{{Starosta {\em et al.}}}},
  \bibinfo{journal}{Nucl.\ Instrum.\ Methods\ Phys.\ Res.\ A,}
  \textbf{\bibinfo{volume}{423}}, \bibinfo{pages}{16} (\bibinfo{year}{1999}).

\bibitem[{\citenamefont{Aoki et~al.}(1979)\citenamefont{Aoki, Furuno, Tagishi,
  Ohya, and Ruan}}]{Aoki}
\bibinfo{author}{\bibfnamefont{T.}~\bibnamefont{Aoki}},
  \bibinfo{author}{\bibfnamefont{K.}~\bibnamefont{Furuno}},
  \bibinfo{author}{\bibfnamefont{Y.}~\bibnamefont{Tagishi}},
  \bibinfo{author}{\bibfnamefont{S.}~\bibnamefont{Ohya}}, \bibnamefont{and}
  \bibinfo{author}{\bibfnamefont{J.}~\bibnamefont{Ruan}},
  \bibinfo{journal}{At.\ Data.\ Nucl.\ Data.\ Tables}
  \textbf{\bibinfo{volume}{23}}, \bibinfo{pages}{349} (\bibinfo{year}{1979}).

\bibitem[{\citenamefont{{Der Mateosian} and Sunyar}(1974)}]{Mateosian}
\bibinfo{author}{\bibfnamefont{E.}~\bibnamefont{{Der Mateosian}}}
  \bibnamefont{and} \bibinfo{author}{\bibfnamefont{A.~W.}
  \bibnamefont{Sunyar}}, \bibinfo{journal}{At.\ Data.\ Nucl.\ Data.\ Tables}
  \textbf{\bibinfo{volume}{13}}, \bibinfo{pages}{391} (\bibinfo{year}{1974}).

\bibitem[{\citenamefont{Palit et~al.}(2000)\citenamefont{Palit, Jain, Joshi,
  Nagaraj, Rao, Chintalapudi, and Ghugre}}]{Palit}
\bibinfo{author}{\bibfnamefont{R.}~\bibnamefont{Palit}},
  \bibinfo{author}{\bibfnamefont{H.~C.} \bibnamefont{Jain}},
  \bibinfo{author}{\bibfnamefont{P.~K.} \bibnamefont{Joshi}},
  \bibinfo{author}{\bibfnamefont{S.}~\bibnamefont{Nagaraj}},
  \bibinfo{author}{\bibfnamefont{B.~V.~T.} \bibnamefont{Rao}},
  \bibinfo{author}{\bibfnamefont{S.~N.} \bibnamefont{Chintalapudi}},
  \bibnamefont{and} \bibinfo{author}{\bibfnamefont{S.~S.}
  \bibnamefont{Ghugre}}, \bibinfo{journal}{Pramana}
  \textbf{\bibinfo{volume}{54}}, \bibinfo{pages}{347} (\bibinfo{year}{2000}).

\bibitem[{\citenamefont{Yamazaki}(1967)}]{Yamazaki}
\bibinfo{author}{\bibfnamefont{T.}~\bibnamefont{Yamazaki}},
  \bibinfo{journal}{Nucl.\ Data.\ A} \textbf{\bibinfo{volume}{3}},
  \bibinfo{pages}{1} (\bibinfo{year}{1967}).

\bibitem[{\citenamefont{{Mason {\em et al.}}}(2005)}]{Mason}
\bibinfo{author}{\bibfnamefont{P.}~\bibnamefont{{Mason {\em et al.}}}},
  \bibinfo{journal}{Phys.\ Rev.\ C} \textbf{\bibinfo{volume}{71}},
  \bibinfo{pages}{014316} (\bibinfo{year}{2005}).

\bibitem[{\citenamefont{Krane and Steffan}(1970)}]{Krane2}
\bibinfo{author}{\bibfnamefont{K.~S.} \bibnamefont{Krane}} \bibnamefont{and}
  \bibinfo{author}{\bibfnamefont{R.~M.} \bibnamefont{Steffan}},
  \bibinfo{journal}{Phys.\ Rev.\ C} \textbf{\bibinfo{volume}{2}},
  \bibinfo{pages}{724} (\bibinfo{year}{1970}).

\bibitem[{NND()}]{NNDC}
\emph{\bibinfo{title}{{NNDC} {O}nline {D}ata {S}ervice}},
  \bibinfo{note}{\url{http://www.nndc.bnl.gov}}.

\bibitem[{\citenamefont{Ajzenberg-Selove
  et~al.}(1977)\citenamefont{Ajzenberg-Selove, Flynn, Orbesen, and
  Sunier}}]{Ajzen}
\bibinfo{author}{\bibfnamefont{F.}~\bibnamefont{Ajzenberg-Selove}},
  \bibinfo{author}{\bibfnamefont{E.~R.} \bibnamefont{Flynn}},
  \bibinfo{author}{\bibfnamefont{S.}~\bibnamefont{Orbesen}}, \bibnamefont{and}
  \bibinfo{author}{\bibfnamefont{J.~W.} \bibnamefont{Sunier}},
  \bibinfo{journal}{Phys.\ Rev.\ C} \textbf{\bibinfo{volume}{15}},
  \bibinfo{pages}{1} (\bibinfo{year}{1977}).

\bibitem[{\citenamefont{Pritychenko et~al.}(2000)\citenamefont{Pritychenko,
  Glasmacher, Brown, Cottle, Ibbotson, Kemper, and Scheit}}]{Pritychenko}
\bibinfo{author}{\bibfnamefont{B.~V.} \bibnamefont{Pritychenko}},
  \bibinfo{author}{\bibfnamefont{T.}~\bibnamefont{Glasmacher}},
  \bibinfo{author}{\bibfnamefont{B.~A.} \bibnamefont{Brown}},
  \bibinfo{author}{\bibfnamefont{P.~D.} \bibnamefont{Cottle}},
  \bibinfo{author}{\bibfnamefont{R.~W.} \bibnamefont{Ibbotson}},
  \bibinfo{author}{\bibfnamefont{K.~W.} \bibnamefont{Kemper}},
  \bibnamefont{and} \bibinfo{author}{\bibfnamefont{H.}~\bibnamefont{Scheit}},
  \bibinfo{journal}{Phys.\ Rev.\ C} \textbf{\bibinfo{volume}{66}},
  \bibinfo{pages}{051601(R)} (\bibinfo{year}{2000}).

\bibitem[{\citenamefont{Skorka et~al.}(1966)\citenamefont{Skorka, Hertel, and
  Retz-Schmidt}}]{Skorka}
\bibinfo{author}{\bibfnamefont{S.~J.} \bibnamefont{Skorka}},
  \bibinfo{author}{\bibfnamefont{J.}~\bibnamefont{Hertel}}, \bibnamefont{and}
  \bibinfo{author}{\bibfnamefont{T.~W.} \bibnamefont{Retz-Schmidt}},
  \bibinfo{journal}{Nucl.\ Data.\ A} \textbf{\bibinfo{volume}{2}},
  \bibinfo{pages}{347} (\bibinfo{year}{1966}).

\bibitem[{\citenamefont{Nixon et~al.}(1975)\citenamefont{Nixon, Jones, Lornie,
  Nagel, Nolan, Price, and Twin}}]{Nixon}
\bibinfo{author}{\bibfnamefont{M.~R.} \bibnamefont{Nixon}},
  \bibinfo{author}{\bibfnamefont{G.~D.} \bibnamefont{Jones}},
  \bibinfo{author}{\bibfnamefont{P.~R.~G.} \bibnamefont{Lornie}},
  \bibinfo{author}{\bibfnamefont{A.}~\bibnamefont{Nagel}},
  \bibinfo{author}{\bibfnamefont{P.~J.} \bibnamefont{Nolan}},
  \bibinfo{author}{\bibfnamefont{H.~G.} \bibnamefont{Price}}, \bibnamefont{and}
  \bibinfo{author}{\bibfnamefont{P.~J.} \bibnamefont{Twin}},
  \bibinfo{journal}{J.\ Phys.\ G:\ Nucl.\ Phys.} \textbf{\bibinfo{volume}{1}},
  \bibinfo{pages}{430} (\bibinfo{year}{1975}).

\bibitem[{\citenamefont{Sterrenburg et~al.}(1997)\citenamefont{Sterrenburg,
  {Van~Middlekoop}, and {Van~Eijkern}}}]{Sterrenburg}
\bibinfo{author}{\bibfnamefont{W.~A.} \bibnamefont{Sterrenburg}},
  \bibinfo{author}{\bibfnamefont{G.}~\bibnamefont{{Van~Middlekoop}}},
  \bibnamefont{and} \bibinfo{author}{\bibfnamefont{F.~E.~H.}
  \bibnamefont{{Van~Eijkern}}}, \bibinfo{journal}{Nucl.\ Phys.\ A}
  \textbf{\bibinfo{volume}{275}}, \bibinfo{pages}{48} (\bibinfo{year}{1997}).

\bibitem[{\citenamefont{Brown and Rae}(2007)}]{NuShell}
\bibinfo{author}{\bibfnamefont{B.~A.} \bibnamefont{Brown}} \bibnamefont{and}
  \bibinfo{author}{\bibfnamefont{W.~D.~M.} \bibnamefont{Rae}},
  \emph{\bibinfo{title}{{MSU}-{NSCL} {R}eport}} (\bibinfo{year}{2007}).

\bibitem[{\citenamefont{Warburton et~al.}(1990)\citenamefont{Warburton, Becker,
  and Brown}}]{Warburton}
\bibinfo{author}{\bibfnamefont{E.~K.} \bibnamefont{Warburton}},
  \bibinfo{author}{\bibfnamefont{J.~A.} \bibnamefont{Becker}},
  \bibnamefont{and} \bibinfo{author}{\bibfnamefont{B.~A.} \bibnamefont{Brown}},
  \bibinfo{journal}{Phys.\ Rev.\ C} \textbf{\bibinfo{volume}{41}},
  \bibinfo{pages}{1147} (\bibinfo{year}{1990}).

\bibitem[{\citenamefont{Brown}()}]{Brown}
\bibinfo{author}{\bibfnamefont{B.~A.} \bibnamefont{Brown}},
  \bibinfo{note}{\url{http://www.nscl.msu.edu/~brown/resources/SDE.HTM#a34t2}}.

\bibitem[{\citenamefont{{Ionescu-Bujor {\em et al.}}}(2006)}]{Bujor}
\bibinfo{author}{\bibfnamefont{M.}~\bibnamefont{{Ionescu-Bujor {\em et al.}}}},
  \bibinfo{journal}{Phys.\ Rev.\ C} \textbf{\bibinfo{volume}{73}},
  \bibinfo{pages}{024310} (\bibinfo{year}{2006}).

\bibitem[{\citenamefont{Wagner et~al.}(1973)\citenamefont{Wagner, Coffin, Ali,
  Alburger, and Gallmann}}]{Wagner}
\bibinfo{author}{\bibfnamefont{P.}~\bibnamefont{Wagner}},
  \bibinfo{author}{\bibfnamefont{J.~P.} \bibnamefont{Coffin}},
  \bibinfo{author}{\bibfnamefont{M.~A.} \bibnamefont{Ali}},
  \bibinfo{author}{\bibfnamefont{D.~E.} \bibnamefont{Alburger}},
  \bibnamefont{and} \bibinfo{author}{\bibfnamefont{A.}~\bibnamefont{Gallmann}},
  \bibinfo{journal}{Phys.\ Rev.\ C} \textbf{\bibinfo{volume}{7}},
  \bibinfo{pages}{2418} (\bibinfo{year}{1973}).

\bibitem[{\citenamefont{Ragan et~al.}(1969)\citenamefont{Ragan, Moss, Poore,
  Roberson, Mitchell, and Tilley}}]{Ragan}
\bibinfo{author}{\bibfnamefont{C.~E.} \bibnamefont{Ragan}},
  \bibinfo{author}{\bibfnamefont{C.~E.} \bibnamefont{Moss}},
  \bibinfo{author}{\bibfnamefont{R.~V.} \bibnamefont{Poore}},
  \bibinfo{author}{\bibfnamefont{N.~R.} \bibnamefont{Roberson}},
  \bibinfo{author}{\bibfnamefont{G.~E.} \bibnamefont{Mitchell}},
  \bibnamefont{and} \bibinfo{author}{\bibfnamefont{D.~R.}
  \bibnamefont{Tilley}}, \bibinfo{journal}{Phys.\ Rev.}
  \textbf{\bibinfo{volume}{188}}, \bibinfo{pages}{1806} (\bibinfo{year}{1969}).

\end{thebibliography}
\clearpage

%----------------------------------------------------------------------------------------------------------

\begin{table}[htbp]
\caption{\label{tab:tableI}Comparison of theoretical and experimental asymmetry
and theoretical and experimental $R_{DCO}$ values in $^{33}$P and $^{33,34}$S}
%\begin{ruledtabular}
\begin{tabular}{cccccc}
\hline
$E_{\gamma}$ & Mixing ratio & $\Delta_{IPDCO}$
& $\Delta_{IPDCO}$ & $R_{DCO}$ & $R_{DCO}$\\
$[$keV$]$ & ($\delta$)from NNDC & $($Experimental$)$ & $($Theoretical$)$ & $($Experimental$)$ & $($Theoretical$)$ \\
\hline
&\multicolumn{4}{c}{$^{34}$S}\\
2127    &   0.0     &  0.053(24)  &  0.043  &            & 1.00        \\
2561    &   0.0     &  0.032(22)  &  0.019  & 0.99(6) & 1.00        \\
1001    &   -0.05   &  0.032(19)  &  0.051  & 0.57(3) & 0.49        \\
1066    &   0.0     &  0.068(25)  &  0.067  & 1.20(7) & 1.20        \\
\hline
&\multicolumn{4}{c}{$^{33}$P}\\
1848    &   -0.03   &  0.050(20)  &  0.045  & 1.76(34) &        \\
1432    &   -0.60   &  -0.022(20) &  -0.002 & 0.89(6) &        \\
416     &   0.09    &  -0.188(29) &  -0.135 & 0.87(6) &        \\
2379    &   0.01    &  0.017(15)  &  -0.014 & 0.48(10) & 0.49   \\
1413    &   0.0     &  0.048(10)  &  0.046  & 0.94(6) & 1.01   \\
\hline
&\multicolumn{4}{c}{$^{33}$S}\\
968    &   -0.02   &  0.032(11)  &  0.047  & 0.84(6) &         \\
1968   &   -0.56   &  0.005(9)  &  0.004  & 1.51(18) &  1.67   \\
\hline
\hline
\end{tabular}
%\end{ruledtabular}
\end{table}

%-------------------------------------------------------------------------

\newpage
\begin{longtable}{cccccccc}
\caption{\label{tableII}The transition energies, excitation energies, relative intensities,
initial and final spin-parities, multipolarities and the $\gamma$-ray
branching ratios in $^{33,34}$P and $^{33}$S.}
\endfirsthead
\caption[]{continued...} \\
    $E_{\gamma}^{\footnotemark[1]}$ &  $E_{x}$ &  $I_{\gamma}$ & $J^\pi_i$   & $J^\pi_f$ &  Multipolarity& \multicolumn{2}{c}{Branching ratios $(\%)$}\\
$[keV]$  & $[keV]$  & $(\%)$   &   &    &   &   Present work & Previous work    \\
\hline
\endhead
\multicolumn{5}{r} {continued...~}\\
\endfoot
\hline
\endlastfoot
$E_{\gamma}^{\footnotemark[1]}$ &  $E_{x}$ &  $I_{\gamma}$  & $J^\pi_i$   & $J^\pi_f$ &  Multipolarity& \multicolumn{2}{c}{Branching ratios $(\%)$}\\
$[keV]$  & $[keV]$  & $(\%)$   &   &    &   &   Present work & Previous work    \\
\hline
$^{33}$P & & & & & & & \\
186.2    &   5640   &   33.2(12)      &  $11/2^{-}$   &  $9/2^{-}$    & (M1)  & 41.0(18)& 46$\pm$2$^{b}$          \\
237.0    &   4227   &   $weak$        &  $7/2^{-}$    &               &       &           &                   \\
247.0    &   5454   &   weak          &  $9/2^{-}$    &               &       &           &                   \\
416.4    &   1848   &   6.95(22)      &  $5/2^{+}_{1}$&  $3/2^{+}$    & M1+E2 & 6.50(28)  & 7$\pm$4$^{a}$          \\
736.2    &   4227   &   8.25(27)      &  $7/2^{-}$    &  $5/2^{+}$    &  (E1) & 7.76(34)  & 11$\pm$2$^{b}$         \\
979.5    &   5207   &   $weak$        &               &  $7/2^{-}$    &       &           &                   \\
993.7    &   5221   &   $weak$        &               &  $7/2^{-}$    &  (D)  &           &                   \\
1008.0   &   5235   &   $weak$        &               &  $7/2^{-}$    &       &           &                   \\
1028.3   &   7966   &   2.18(8)       &  $17/2^{(+)}$ &  $15/2^{(-)}$ &  (E1) &           &                   \\
1226.8   &   5454   &   50.2(16)      &  $9/2^{-}$    &  $ 7/2^{-}$   &  (M1) & 91.5(40)  & 100$^{b}$         \\
1298.1   &   6938   &   5.45(19)      &  $15/2^{(-)}$ &  $11/2^{-}$   &  (E2) &           &                   \\
1311.5   &   6952   &   $weak$        &               &  $11/2^{-}$   &  (D)  &           &                   \\
1412.6   &   5640   &   47.8(16)      &  $11/2^{-}$   &  $7/2^{-}$    & E2+M3 & 59.0(24)& 54$\pm$2$^{b}$         \\
1432.1   &   1432   &   $>$ 18.95     &  $3/2^{+}$    &  $1/2^{+}$    & M1+E2 & 100       & 100$^{a}$         \\
1642.7   &   3491   &   6.93(25)      &  $5/2^{+}_{2}$&  $5/2^{+}_{1}$&  Q    & 54.1(24)  & 62$\pm$2$^{a}$         \\
1780.5   &   3629   &   3.86(15)      &  $7/2^{+}$    &  $5/2^{+}_{1}$&  (M1) & 36.9(17)  & 28$\pm$3$^{a}$         \\
1825.1   &   5454   &   4.68(15)      &  $9/2^{-}$    &  $7/2^{+}$    &   D   & 8.5(37)   &                   \\
1848.1   &   1848   &   100(3)        &  $5/2^{+}_{1}$&  $1/2^{+}$    & E2+M3 & 93.5(39)  & 93$\pm$4$^{a}$         \\
2058.8   &   3491   &   5.25(19)      &  $5/2^{+}_{2}$&  $3/2^{+}$    & (M1)  & 41.0(18)  & 38$\pm$4$^{a}$         \\
2141.6   &   3990   &   1.45(7)       &               &  $5/2^{+}_{1}$& (Q)   &           &                   \\
2196.5   &   3629   &   6.60(21)      &  $7/2^{+}$    &  $3/2^{+}$    & (E2)  & 63.1(25)  & 72$\pm$3$^{a}$         \\
2378.8   &   4227   &   98.0(30)      &  $7/2^{-}$    &  $5/2^{+}_{1}$& E1+M2 & 92.2(39)  & 89$\pm$2$^{a}$         \\
3606.0   &   5454   &   $weak$        &  $9/2^{-}$    &  $5/2^{+}_{1}$& (M2)  &           &                   \\
3491.1   &   3491   &   0.62(3)       &  $5/2^{+}_{2}$&  $1/2^{+}$    & (E2)  & 4.84(26)  & $<$ 4$^{a}$               \\
\hline
\hline
$^{34}$P & & & & & & & \\
429.4    &   429    &   100(3)        &  $2^{+}$      &  $1^{+}$      & M1+E2   & 100        &                     \\
679.4    &   4630   &   3.17(16)      &  $6^{(-)}$    &  $5^{(-)}$    & (M1)    & 42.9(31)   &                    \\
1047.8   &   3353   &   17.10(91)     &  $5^{(-)}$    &  $4^{(-)}$    & (M1)    &            &                     \\
1179.8   &   1609   &   0.74(10)      &  $1^{(+)}$    &  $2^{+}$      & (M1)    &            &                     \\
1443.7   &   3749   &   $weak$        &               &  $4^{(-)}$    &         &            &                        \\
1607.1   &   6237   &   4.50(45)      &  $7^{(+)}$    &  $6^{(-)}$    & (E1)    & 40.5(41) &                     \\
1646.2   &   3951   &   3.60(43)      &  $5^{(-)}$    &  $4^{(-)}$    & (M1)    &            &                    \\
1637.7   &   3943   &   $weak$        &               &  $4^{(-)}$    &         &            &                    \\
1876.1   &   2305   &   88.6(11)      &  $4^{(-)}$    &  $2^{+}$      & (M2+E3) & 100        &                    \\
1891.6   &   2321   &   11.28(88)     &  $(3^{-})$    &  $2^{+}$      & (E1+M2) &            &                     \\
2325.1   &   4630   &   4.22(30)      &  $6^{(-)}$    &  $4^{(-)}$    &  (E2)   & 57.1(48)   &                    \\
2884.3   &   6237   &   6.61(40)      &  $7^{(+)}$    &  $5^{(-)}$    &  (M2)   & 59.5(48)   &                    \\
3931.7   &   6237   &   $weak$        &  $7^{(+)}$    &  $4^{(-)}$    &         &            &                    \\
\hline
\hline
$^{33}$S & & & & & & & \\
597.0    &   5990   &                 &               &              &        &           &                     \\
597.0    &   5393   &                 &               &  $(11/2^{-})$&        &           &                      \\
602.5    &   3539   &   0.11(1)       &               &  $7/2^{-}$   &        &           &                      \\
841.1    &   841    &   $>$ 0.63      &  $1/2^{+}$    &  $3/2^{+}$   & (M1+E2)&           &                       \\
845.1    &   3781   &   0.67(3)       &  $(9/2^{+})$  &  $7/2^{-}$   & (E1)   &           &                      \\
968.4    &   2936   &   48.5(17)      &  $7/2^{-}$    &  $5/2^{+}$   & E1+M2  & 51.0(23)  & 50$^{c}$              \\
1015.3   &   4796   &   0.37(2)       &  $(11/2^{-})$ &  $(9/2^{+})$ & (E1)   &           &                       \\
1126.5   &   1968   &   0.63(2)       &  $5/2^{+}$    &  $1/2^{+}$   & (E2)   & 0.63(3)   & $<$ 1.5$^{c}$            \\
1931.2   &   4867   &   0.62(3)       &               &  $7/2^{-}$   &        &           &                      \\
1967.6   &   1968   &   100(3)        &  $5/2^{+}$    &  $3/2^{+}$   & M1+E2  & 99.4(42)  & 100$^{c}$             \\
2935.6   &   2936   &   46.7(16)      &  $7/2^{-}$    &  $3/2^{+}$   & (M2+E3)& 49.0(27)  & 50$^{c}$              \\
\hline
\end{longtable}
\footnotetext[1] { The quoted energies are within $\pm$1 keV.}
\footnotetext[2] { $^{a}$Ref.~\cite{Wagner}}
\footnotetext[3] { $^{b}$Ref.~\cite{Nixon}}
\footnotetext[4] { $^{c}$Ref.~\cite{Ragan}}

%-------------------------------------------------------------------------------------------------------------------

\newpage
\begin{table}[htbp]
\caption{\label{tab:tableIII}Reduced transition probabilities
for 1876 keV transition in $^{34}$P considering both M2/E3
mixing and E2/M3 mixing. The lifetime range has been taken
from the reported values of Asai {\em et al.}~\cite{Asai}. The mixing ratios
are from our polarization measurements.}
\begin{tabular}{ccccccccccccccc}
\hline
Half-life &   &   & Mixing ratio  &   &  &   & \multicolumn{8}{c}{Reduced transition probabilities}\\
$($ns$)$  &   &   &  ($\delta$)   &   &  &   &\multicolumn{3}{c}{M2/E3} &   &  & \multicolumn{3}{c}{E2/M3}   \\
          &   &   &               &   &  &   & $B(M2)$  &  &  $B(E3)$   &   &  &  $B(E2)$  &  $B(M3)$\\
          &   &   &               &   &  &   & $(W.u.)$ &  &  $(W.u.)$  &   &  & $(W.u.)$  & $(W.u.)$\\
\hline
          &   &   &               &   &  &   &          &  &            &   &  &           &         \\
          &   &   &  -1.03        &   &  &   & 0.207    &  &  372.681   &   &  &   0.006    &  12520.654 \\[-1ex]
\raisebox{1.5ex}{0.3}    &   &   & -0.27   &   &  &   & 0.397    &  &  49.191   &   &  &   0.012    &  1652.638        \\
          &   &   &  -1.03        &   &  &   & 0.025    &  &  44.722   &   &  &   0.001    &  1502.478 \\[-1ex]
\raisebox{1.5ex}{2.5}    &   &   & -0.27   &   &  &   & 0.048    &  &  5.903   &   &  &   0.001    &  198.317        \\
\hline
\hline
\end{tabular}
%\end{ruledtabular}
\end{table}

%-----------------------------------------------------------------------------------------------------------------

\newpage
\begin{table}[htbp]
\caption{\label{tab:tableIV}Comparison between experimental and theoretical transition
energies, excitation energies, mixing ratios and reduced transition probabilities
in $^{35}$S and $^{37}$Ar.}
\begin{ruledtabular}
\begin{tabular}{cccccc}
$E_\gamma$[keV] & $E_x(J^{\pi})[keV]$ & $\tau$[ns] & $\delta$ & {\em B(M2)}~[W.u.] & {\em B(E3)}~[W.u.]\\
\multicolumn{1}{c}{Expt.~~~~Theo.} & \multicolumn{1}{c}{Expt.~~~~Theo.} & from NNDC &\multicolumn{1}{c}{Expt.~~~~Theo.}&\multicolumn{1}{c}{Expt.~~~~Theo.} &\multicolumn{1}{c}{Expt.~~~~Theo.}\\
\hline
$^{35}$S & & & & & \\
%& & & & & \\
1911~~~~2738 &1911~~~~2738 & 1.02(5) & -0.19(8)~~~~-0.05 & 0.088(5)~~~~0.196 & 4.62~~~~0.38 \\
\hline
$^{37}$Ar & & & & & \\
%& & & & & \\
1611~~~~2680 &1611~~~~2680 & 4.37(9) & -0.12(1)~~~~-0.08 & 0.058(13)~~~~0.112 & 1.7(3)~~~~0.50 \\
\end{tabular}
\end{ruledtabular}
\end{table}

%---------------------------------------------------------------------------------------------------------------

%--------------------------------------------------------------------------

%---------------------------------------------------------------------------

%
%%---------------------------------------------------------------------------
%

%--------------------------------------------------------------------------

%--------------------------------------------------------------------------

%
%----------------------------------------------------------------------------

%---------------------------------------------------------------------------

%----------------------------------------------------------------------------

%---------------------------------------------------------------------------

%--------------------------------------------------------------------------

%----------------------------------------------------------------------

%-----------------------------------------------------------------------

%------------------------------------------------------------------------

%------------------------------------------------------------------------

%-----------------------------------------------------------------------

%---------------------------------------------------------------------

\clearpage

%---------------------------------------------------------------------

%---------------------------------------------------------------------------
\clearpage
%----------------------------------------------------------------------------

%----------------------------------------------------------------------------

%\bibliography{draft3}
\clearpage
\end{document}